\documentclass[twocolumn,amsmath,amssymb,10pt,superscriptaddress,a5paper,letterpaper,fleqn,nofootinbib,preprintnumbers]{revtex4-1}
\usepackage{amssymb}
\usepackage{epsfig}
\usepackage{graphicx}
\usepackage{dcolumn}
\usepackage{array}
\usepackage{bm}
\usepackage{fancyhdr}
\usepackage{longtable}
\usepackage{multirow}
\usepackage{float}
\usepackage{afterpage}  
\usepackage{color}
\usepackage{tabularx}

\usepackage[warnundef]{jabbrv}
\DefineJournalAbbreviation{Test}{Tes} 
\setlongtables
\newcommand {\nc} {\newcommand}
\nc {\IR} [1] {\textcolor{red}{#1}} 
\nc {\IB} [1] {\textcolor{blue}{#1}}
\nc {\IC} [1] {\textcolor{cyan}{#1}}

\nc {\cgmf} {$\texttt{CGMF}$}
\nc {\beoh} {$\texttt{BeoH}$}

\nc {\grays} {$\gamma$ rays}
\nc {\gray}  {$\gamma$-ray}

\begin{document}
\setcounter{page}{1}
\title{
     \qquad \\ \qquad \\ \qquad \\  \qquad \\  \qquad \\ \qquad \\ 
New Procedure for the Evaluation of Fission Product Yields:  \\ Application to the Spontaneous Fission of $^{252}$Cf
}

\author{A.E.~Lovell}
   \email[Corresponding author: ]{lovell@lanl.gov}
   \affiliation{Los Alamos National Laboratory, Los Alamos, NM 87545, USA}
\author{T.~Kawano} 
   \affiliation{Los Alamos National Laboratory, Los Alamos, NM 87545, USA} 
\author{P.~Talou} 
   \affiliation{Los Alamos National Laboratory, Los Alamos, NM 87545, USA} 
   \affiliation{Stardust Science Labs, Santa Fe, NM 87507, USA}

\date{\today}
\preprint{LA-UR-26-22230}
   \received{xx July 2018; revised received xx September 2018; accepted xx October 2018}

\begin{abstract}{
Over the last decade, there has been significant improvement in the understanding and modeling of the decay of fission fragments by both prompt and delayed emission.  These model improvements open the door for performing consistent evaluations across multiple fission observables, providing not only mean values but also covariances between observables.  One such model is the Hauser-Feshbach Fission Fragment Decay model implemented in \beoh{}, which uses distributions of initial conditions of fission fragments to perform a Hauser-Feshbach decay for prompt neutron and $\gamma$-ray emission and evaluated decay data to calculate cumulative fission product yields.  This manuscript describes a new evaluation procedure for independent and cumulative fission product yields, including full correlations among the fission products.  We use a Bayesian Kalman filter to fit both experimental cumulative fission product yields and those from the ENDF/B-VIII.0 evaluated library, producing mean values and covariances.  In addition to comparing the fission products from these optimizations, we calculate prompt and delayed neutron and $\gamma$-ray multiplicities using the fitted parameters and compare to some available experimental data.  We see reasonable agreement, even when these quantities are not included in the optimization.
}
\end{abstract}
\maketitle

\lhead{Towards an Updated $\dots$}          
\chead{NUCLEAR DATA SHEETS}                             
\rhead{A.E.~Lovell \textit{et al.}}                   
\lfoot{}                                                           
\rfoot{}                                                          
\renewcommand{\footrulewidth}{0.4pt}
\tableofcontents{}


\section{INTRODUCTION}
\label{sec:intro}

In a typical nuclear fission reaction, hundreds of fission fragments are produced in a variety of configurations characterized by their mass, $A$, charge, $Z$, kinetic energy, KE, excitation energy, $U$, spin, $J$, and parity $\pi$. Those fission fragments emit a number of prompt neutrons and $\gamma$ rays to reach their ground state or a long-lived isomeric state. The distribution of these fission products in mass and charge, $Y_I(A,Z)$, is called the Independent Fission Yields (IFYs). Those fission products will further $\beta$ decay and possibly emit another round of $\beta$-delayed neutrons and $\gamma$ rays to reach their final stable state. Their final distribution in $A$ and $Z$, $Y_C(A,Z)$, is called Cumulative Fission Yields (CFYs). The accurate knowledge of fission yields is important for fundamental studies of the dynamical fission process and for a range of applications and in particular at every stage of the nuclear fuel cycle. Our current knowledge of IFYs and CFYs for a number of isotopes and fission reactions is captured in existing evaluated nuclear data libraries such as the US ENDF/B-VIII.1 \cite{ENDFB81arxiv}, the Japanese JENDL-5 \cite{JENDL5}, and the OECD (Organisation for Economic Co-Operation and Development) JEFF-4.0 \cite{JEFF40} libraries.

The latest release of the US library, ENDF/B-VIII.1, contains IFYs and CFYs for 40 isotopes and fission reactions at three incident neutron energies (in addition to spontaneous fission reactions) at thermal, "fast" and 14 MeV.  For ENDF/B-VII.1 \cite{ENDFB71}, an additional neutron energy point at 2.0 MeV was added to the $^{239}$Pu(n,f) evaluated reaction to account for an important linear dependence uncovered to explain a long-standing problem in our understanding of the stockpile \cite{Chadwick2010fpy}. For the other three energies, the ENDF/B evaluations have remained largely unchanged from the England-Rider evaluation work of 1994 who developed fission yield systematics based on experimental data only \cite{England1994}.  For ENDF/B-VIII.1, several minor corrections were performed \cite{Mattera2021,Mattera2023}, including the correction of 17 anomalously large CFY uncertainties near stability, the removal of isomeric states in $^{109}$Ru and $^{109}$Rh that have not been confirmed experimentally, and the correction to thermal neutron-induced fission products of $^{241}$Pu that have been carried through several of the previous releases.  

Here we present an important breakthrough in the evaluation of IFY and CFY through the combination of nuclear reaction model calculations and experimental data, not only on fission yields but also on other fission quantities obtained simultaneously. This new approach provides unique and original results: (1) Fission yields as a function of incident neutron energy on a fine energy grid; (2) The evaluated yields are constrained consistently by experimental data on auxiliary fission quantities such as prompt neutron and $\gamma$-ray multiplicities; and (3) Covariance matrices are obtained for all energy-dependent fission yields as well as cross-energy correlations.

This new approach is presented in detail in this paper for the case of the spontaneous fission of $^{252}$Cf. Although in this case there is no incident neutron energy dependence, the theory, methods and codes can be similarly applied to neutron-induced fission reactions up to 20 MeV, as demonstrated in \cite{Lovell2023,Lovell2025}.

Section \ref{sec:theory} presents the \beoh{} code that models the de-excitation of scission fragments through prompt and $\beta$-delayed neutron and $\gamma$-ray emission. Relationships between IFY, CFY, and prompt and delayed neutron multiplicities are explained. Section \ref{sec:fitting} introduces the Bayesian Kalman filter technique that is used to optimize the \beoh{} model parameters to best reproduce experimental data. Sensitivity calculations are performed for the relevant model parameters, and the experimental data selected for the optimization phase of the Bayesian updating procedure are discussed. Section \ref{sec:results} presents the results obtained when fitting to either experimental data or to ENDF/B values. Finally, the correlation and covariance matrices obtained for IFY and CFY are analyzed in Sec. \ref{sec:covariance}.

\section{THEORY}
\label{sec:theory}


\beoh{} is a deterministic fission fragment decay code \cite{Okumura2018,Lovell2021} that takes information about the fissioning compound system for a given incident neutron energy, constructs the yields of the scission fragments, then follows their decay through prompt neutron and \gray\ emission using the Hauser-Feshbach statistical theory \cite{HauserFeshbach} and through delayed neutron and \gray\ emission using a time-independent formulation.  We give an overview of the model here, but a more detailed description can be found in \cite{Lovell2021}. For the case of spontaneous fission studied in this paper, the calculation is simpler, as no energy dependence in the entrance channel has to be considered. To compute the emission of neutrons and \grays\ using the Hauser-Feshbach statistical theory, the initial conditions of the fission fragments right after scission are needed, namely their yields in mass, charge, total kinetic energy, spin, and parity, $Y(A,Z,\mathrm{TKE},J,\pi)$. 

The mass distribution, $Y(A)$, is constructed as a sum of Gaussian distributions,
\begin{equation}
    Y(A) = G_0(A) + G_1(A) + G_2(A)
\end{equation}
where
\begin{equation}
\label{eqn:symmMode}
    G_0 = \frac{W_0}{\sqrt{2\pi} \sigma_0} \exp \left [ -\frac{(A-A_c/2)^2}{2\sigma_0 ^2} \right ]
\end{equation}
and 
\begin{eqnarray}
    \nonumber G_{i=1,2} & = & \frac{W_i}{\sqrt{2\pi} \sigma_i} \left \{ \exp \left [ -\frac{(A-\mu_i)^2}{2\sigma_i ^2} \right ] \right . \\
   & + &  \left . \exp \left [ -\frac{(A-[A_c-\mu_i])^2}{2\sigma_i ^2} \right ] \right \}.
\end{eqnarray}
Here, the mass of the compound fissioning nucleus, $A_c$, is equal to 252.  To keep consistency with the energy-dependent parametrizations, we define the weights as
\begin{equation}
    W_{i=1,2} = \frac{1}{1+\exp (w_i^a/w_i^b)},
\end{equation}
although, clearly, $w_i^a$ and $w_i^b$ are not independent in this case.  The weight of the symmetric mode, $W_0$, is defined through the normalization condition, $2=W_0+2W_1+2W_2$. The mean of the symmetric mode is $A_c/2$, as in Eq. (\ref{eqn:symmMode}).  The mass-dependent charge distribution, $Y(Z|A)$, is taken from Wahl systematics \cite{Wahl2002}, and we additionally allow scaling factors, $f_Z$ and $f_N$, 
\begin{equation}
    F_Z = 1 + (F_Z^W + 1)f_Z,
\end{equation}
and
\begin{equation}
    F_N = 1 + (F_N^W + 1)f_N,
\end{equation}
where $F_Z^W$ and $F_N^W$ are the original parameters from the Wahl systematics that govern the odd-even effect in $Y(Z|A)$.  These factors have been found to be important in reproducing the average delayed neutron multiplicity, $\overline{\nu}_d$ \cite{Okumura2022}.

The magnitude of the total kinetic energy, TKE, is given as an input to \beoh{}, and fitted in this work to available experimental data (shown later).  The fragment mass dependence of the average total kinetic energy, $\overline{\rm TKE}(A)$, and its width, $\sigma_\mathrm{TKE}(A)$, are defined using the same functional form as in Ref.~\cite{Lovell2021}.  To reduce the number of free parameters included in the fit, the parameters in these two functions are held constant. The TKE is split between the fission fragment pairs based on kinematics.  

The total excitation energy, TXE, is calculated from the Q-value of each fragment split, $\mathrm{TXE} = Q - \mathrm{TKE}$.  Then, the TXE is shared between the two fragments based on a ratio of temperatures of the fragments
\begin{equation}
\label{eqn:RTA}
    R_T(A) = \frac{T_l}{T_h} = \sqrt{\frac{U_l a_h(U_h)}{U_h a_l(U_l)}},
\end{equation}
where $l$ and $h$ represent the light and heavy fragments respectively, $(U)$ is the energy-dependent level density parameter from the Gilbert-Cameron level density model \cite{Gilbert1965}, and $U$ is the excitation energy corrected by the pairing energy, $U = E_x-\Delta$.  The mass dependence of $R_T(A)$ is included to better reproduce experimental $\overline{\nu}_p(A)$, the average prompt neutron multiplicity for the mass $A$ of the fragment.  

A joint distribution over excitation energy, $E_x$, spin, $J$, and parity, $\pi$, is created for the initial population of the light and heavy fragments, $P_{l,h}(E_x,J,\pi)$,
\begin{equation}
\label{eqn:pop}
    P_{l,h}(E_x,J,\pi) = \frac{1}{2} G_{l,h}(E_x) R_{l,h}(J),
\end{equation}
where positive and negative spins are assumed to be equally probable, leading to the factor of 1/2. The excitation energy distribution is assumed to be Gaussian,
\begin{equation}
\label{eqn:Exdist}
    G_{l,h}(E_x) = \frac{1}{\sqrt{2\pi}\delta _{l,h}} \exp \left \{ - \frac{(E_x-E_{l,h})^2}{2\delta^2_{l,h}} \right \}
\end{equation}
with
\begin{equation}
\label{eqn:Jdist}
    R_{l,h}(J) = \frac{J+1/2}{f^2\sigma_{l,h}^2(U)} \exp \left \{ - \frac{(J+1/2)^2}{2f^2\sigma^2_{l,h}(U)} \right \}.
\end{equation}
In Eq. (\ref{eqn:Exdist}), the width, $\delta_{l,h}$, is proportional to the width of the mass-dependent total kinetic energy distribution.  The spin cutoff factor, $\sigma^2_{l,h}(U)$, in Eq. (\ref{eqn:Jdist}) additionally is multiplied by an adjustable scaling factor, $f$.  The population distributions, $P_{l,h}$ of Eq. (\ref{eqn:pop}), are first created for the heavy and light fragments.  Then the statistical Hauser-Feshbach decay is performed for each fragment, using a discretized continuum, where the highest energy bin is $U_{l,h}+2\delta_{l,h}$.  The population of each residual nucleus is tracked, along with results for the spectra of emitted neutrons and $\gamma$ rays.  

Because Eq. (\ref{eqn:pop}) is normalized to unity, the prompt neutron multiplicity for each fragment, $\overline{\nu}_{l,h}$, is calculated from the neutron spectra for that fragment, $\phi _{l,h}(\epsilon)$, as
\begin{equation}
    \int \phi _{l,h}(\epsilon) d\epsilon = \overline{\nu}_{l,h}.
\end{equation}
The average total prompt neutron multiplicity for each post-scission fission fragment pair, labeled by $k$, is given by $\overline{\nu}_k= \overline{\nu}_l + \overline{\nu}_h$.  To calculate $\overline{\nu}_p$, each $\overline{\nu}_k$ is weighted by its pre-emission yield and summed.

The independent fission product yields, $Y_I(A,Z,M)$, where the index $M$ is now included to keep track of isomeric states, are, by definition, the yields after prompt neutron emission.  The cumulative fission product yields, i.e., after $\beta$ decay and $\beta$-delayed neutron emission, are calculated in a time-independent framework in \beoh{} using decay data from the ENDF/B-VIII.0 decay data library\footnote{We note that this work was performed before the release of the ENDF/B-VIII.1 library but follow-on studies are using the updated decay data} \cite{ENDFB8},
\begin{eqnarray}
    \nonumber Y_C(A_i,Z_i,M_i) = Y_I(A_i,Z_i,M_i) \\
    + \sum \limits _{j=1} ^N \sum \limits _{l=1} ^{L_j} Y_C(A_j,Z_j,M_j) b_{jl} \delta _{jl,i}. 
\end{eqnarray}
Here, $N$ is the total number of nuclei that produce the $i$-th nucleus, $(A_i,Z_i,M_i)$, and $\delta _{jl,i}$ connects $Y_C(A_j,Z_j,M_j)$ to $Y_C(A_i,Z_i,M_i)$ through any of the branching ratios, $b_{jl}$, among $L_j$ total decay modes.  As mentioned in the introduction, the use of this type of consistent fission modeling provides for the consistent calculation of all post-scission observables, allowing, for example, the inclusion of prompt observables to constrain delayed ones.

\section{OPTIMIZATION PROCEDURE}
\label{sec:fitting}

\subsection{Kalman Filter}
\label{sec:kalman}


To perform the optimization, we use a type of Bayesian optimization, called a Kalman filter \cite{Kalman1960}, in which the likelihood and prior distribution are assumed to be multivariate Gaussian distributions (see, e.g. \cite{Rising2013} for details).  In the linear approximation, it is assumed that the relationship between the model parameters and the model output is linear,
\begin{equation}
\label{eq:Kalman}
    f(\textbf{x}_1) = f(\textbf{x}_0) + \mathbb{C} \delta \textbf{x},
\end{equation}
where $\textbf{x}$ is the vector of model parameters, $f(\textbf{x})$ is the model, $\mathbb{C}$ is the parameter sensitivity matrix, and $\delta \textbf{x} = \textbf{x}_1 - \textbf{x}_0$ is the difference between the updated and initial parameters. The sensitivity matrix is calculated for each $f_i(\textbf{x}) = $CFY$_i$ from \beoh{},
\begin{equation}
\label{eq:pCov}
    \left . \mathbb{C}_{ij} = \frac{\partial f_i(\textbf{x})}{\partial x_j} \right |_{\textbf{x}=\textbf{x}_{0}},
\end{equation}
where the model difference, $\partial f(\textbf{x})$, is calculated for each $i$ cumulative fission product yield with a change in parameter $x_j$.   

The parameter changes in Eq. (\ref{eq:Kalman}), $\delta \textbf{x} = \textbf{x}_1 - \textbf{x}_0$, are calculated as
\begin{equation}
    \delta \textbf{x} = \mathbb{P} \mathbb{C}^T \mathbb{V}^{-1} (\phi - f(\textbf{x}_0)),
\end{equation}
where $\phi$ is the vector of experimental data to be fitted and the resulting parameter correlation matrix, $\mathbb{P}$, is calculated as
\begin{equation}
    \mathbb{P} = (\mathbb{X}^{-1} + \mathbb{C}^T \mathbb{V}^{-1} \mathbb{C}) ^{-1}.
\end{equation}
To calculate $\mathbb{P}$, the parameter and experimental covariances are required.  The parameter covariance, $\mathbb{X}$, contains information about the prior parameter uncertainty.  The diagonal elements of $\mathbb{V}$ contain the experimental variances or the square of the experimental standard deviations, and the off-diagonal elements contain the experimental correlations in energy.  In this work, we only consider the experimental standard deviations, as reported in literature. However, it is important to note that much work has been performed on the importance of including realistic experimental uncertainties from all sources in optimization, e.g. \cite{Neudecker2020}. Templates of expected uncertainties \cite{Neudecker2023,FYtemplate} have been developed to help identify and include potentially missing sources of experimental uncertainties. 

Once the updated parameters and their covariance matrix are obtained, we can calculate the linear model update using Eq. (\ref{eq:Kalman}).  It is important to compare these updates to the full model calculation, $f(\textbf{x}_1)$, as complex models such as the one described here can be very non-linear.  The observable covariance matrix, $\mathbb{F}$, can also be computed through
\begin{equation}
    \label{eqn:fCov}
    \mathbb{F} = \mathbb{C} \mathbb{P} \mathbb{C}^T.
\end{equation}
The standard deviations for the CFYs are given as the square root of the diagonal elements of $\mathbb{F}$.  Typically, uncertainties extracted directly from the Kalman fitting procedure are unreasonably small, although these values become more realistic when complete experimental uncertainties are taken into account, e.g. \cite{Neudecker2020}.  The covariance matrix is then often scaled by some factor related to the $\chi^2$ of the optimized calculation against the spread of experimental data.  In this work, we scale the covariance matrix by $\chi_i^2/N_i$ for each CFY$_i$ included in the optimization, where $N_i$ is the total number of data values for that fission product that were included in the fit.  For isotopes that were not included in the fit, e.g., isomeric states, too low cumulative yields, or no available experimental data, we scale the calculated covariance matrix by $\chi_\mathrm{tot}^2/N_\mathrm{tot}$, the average $\chi^2$ of all fission product data included in the optimization.  

It is important to check the validity of the linear approximation by using the $\textbf{x}_1$ parameter set in the full \beoh{} calculation and comparing these results to those of the Kalman filter, Eq. (\ref{eq:Kalman}).  We quantify this difference by calculating the percent difference between the two calculations for each cumulative fission product yield
\begin{equation}
\label{eqn:KalmanComp}
    \delta_{KB}(A,Z) =  100 \times \left | \frac{Y_C^{BeoH}(A,Z)-Y_C^{Kalman}(A,Z)}{Y_C^{BeoH}(A,Z)} \right |.
\end{equation}

\subsection{Numerical Details}


The initial parameter values for the fission fragment initial conditions are shown in column 2 of Table \ref{tab:parameters} and are taken from our companion Monte Carlo code, \cgmf{} \cite{CGMF}, which implements similar physics models as \beoh{} but does not include the decay from IFY to CFY.  The scaling for the spin cutoff factor, here $f$, is defined differently between the two codes, so for \beoh{}, $f$ was initially adjusted to reproduce the evaluated $\overline{\nu}_p$.  Although we have a parametrization for TKE$(A)$ and $\sigma_\mathrm{TKE}(A)$, in this work, we only include the magnitude of the average TKE as a free parameter in the optimization.  We also allow the mass-dependent $R_T$ parameters to vary.  Although doing this adds a significant number of parameters, each cumulative fission product yield is sensitive to only a few of these values, as will be shown below.  

\begin{table*}[h]
    \centering
    \caption{Initial \beoh{} parameter values (column two) and fitted parameter values when experimental data (ENDF/B-VIII.0 evaluation) was used in the optimization, column three (five) for the parameters listed in column one.  Column four (six) give the percent changes from the initial parameters.  Details of the parametrizations are given in the text.  TKE is given in MeV; all other parameters are dimensionless.}
    \begin{tabular}{c|ccccc}
    \textbf{Parameter} & \textbf{Initial} & \textbf{Data fit} & \textbf{$\Delta p ^\mathrm{Data}$ (\%)} & \textbf{ENDF fit} & \textbf{$\Delta p ^\mathrm{ENDF}$ (\%)} \\ \hline \hline
    $w_1^a$ & -1.0000 & -1.0048 & 0.4800 & -1.0004 & 0.0400 \\
    $w_1^b$ & -1.7706 & -1.7620 & 0.4857 & -1.7697 & 0.0508 \\
    $\mu_1$ & 15.78 & 15.56 & 1.39 & 15.83 & 0.32 \\
    $\sigma_1$ & 5.7679 & 6.1391 & 6.4356 & 5.9830 & 3.7293 \\
    $w_2^a$ & -1.0000 & -0.9811 & 1.8900 & -1.0071 & 0.7100 \\
    $w_2^b$ & 1.5706 & 1.5595 & 0.7067 & 1.5594 & 0.7131 \\
    $\mu_2$ & 22.026 & 20.2140 & 8.2266 & 20.5592 & 6.6594 \\
    $\sigma_2$ & 7.9062 & 8.0272 & 1.5304 & 7.8835 & 0.2871 \\
    $\sigma_0$ & 10.0496 & 9.9307 & 1.1831 & 9.9986 & 0.5075 \\
    TKE & 185.78 & 185.97 & 0.10 & 185.66 & 0.06 \\
    $R_T(A=126)$ & 1.000 & 1.000 & 0.000 & 1.000 & 0.000 \\
    $R_T(A=127)$ & 1.173 & 1.173 & 0.000 & 1.173 & 0.000 \\
    $R_T(A=128)$ & 1.346 & 1.346 & 0.000 & 1.346 & 0.000 \\
    $R_T(A=129)$ & 1.519 & 1.512 & 0.461 & 1.512 & 0.461 \\
    $R_T(A=130)$ & 1.692 & 1.713 & 1.241 & 1.713 & 1.241 \\
    $R_T(A=131)$ & 1.613 & 1.666 & 3.286 & 1.728 & 7.13 \\
    $R_T(A=132)$ & 1.534 & 1.530 & 0.261 & 1.512 & 1.434 \\
    $R_T(A=133)$ & 1.455 & 1.428 & 1.856 & 1.460 & 0.344 \\
    $R_T(A=134)$ & 1.376 & 1.360 & 1.163 & 1.335 & 2.985 \\
    $R_T(A=135)$ & 1.297 & 1.336 & 3.007 & 1.227 & 5.397 \\
    $R_T(A=136)$ & 1.307 & 1.306 & 0.077 & 1.276 & 2.372 \\
    $R_T(A=137)$ & 1.317 & 1.237 & 6.074 & 1.245 & 5.467 \\
    $R_T(A=138)$ & 1.328 & 1.379 & 3.840 & 1.387 & 4.443 \\
    $R_T(A=139)$ & 1.338 & 1.452 & 8.520 & 1.407 & 5.157 \\
    $R_T(A=140)$ & 1.348 & 1.296 & 3.858 & 1.342 & 0.445 \\
    $R_T(A=141)$ & 1.359 & 1.340 & 1.398 & 1.369 & 0.736 \\
    $R_T(A=142)$ & 1.328 & 1.250 & 5.873 & 1.280 & 3.614 \\
    $R_T(A=143)$ & 1.298 & 1.361 & 4.854 & 1.352 & 4.160 \\
    $R_T(A=144)$ & 1.267 & 1.266 & 0.079 & 1.286 & 1.500 \\
    $R_T(A=145)$ & 1.237 & 1.231 & 0.485 & 1.229 & 0.647 \\
    $R_T(A=146)$ & 1.196 & 1.191 & 0.418 & 1.235 & 3.261 \\
    $R_T(A=147)$ & 1.166 & 1.120 & 3.945 & 1.177 & 0.943 \\
    $R_T(A=148)$ & 1.146 & 1.229 & 7.243 & 1.165 & 1.658 \\
    $R_T(A=149)$ & 1.115 & 1.234 & 10.673 & 1.130 & 1.345 \\
    $R_T(A=150)$ & 1.085 & 1.026 & 5.438 & 1.001 & 7.742 \\
    $R_T(A=151)$ & 1.054 & 0.972 & 7.780 & 1.018 & 3.416 \\
    $R_T(A=152)$ & 1.051 & 0.998 & 5.043 & 1.039 & 1.142 \\
    $R_T(A=153)$ & 0.985 & 0.929 & 5.685 & 0.950 & 3.553 \\
    $R_T(A=154)$ & 0.980 & 1.083 & 10.510 & 1.015 & 3.571 \\
    $R_T(A=155)$ & 0.953 & 0.930 & 2.413 & 0.972 & 1.994 \\
    $R_T(A=156)$ & 0.977 & 0.947 & 3.071 & 0.944 & 3.378 \\
    $R_T(A=157)$ & 0.991 & 0.987 & 0.404 & 0.999 & 0.807 \\
    $R_T(A=158)$ & 0.996 & 1.005 & 0.904 & 1.009 & 1.305 \\
    $R_T(A=159)$ & 1.010 & 1.024 & 1.386 & 1.016 & 0.594 \\
    $R_T(A=160)$ & 1.024 & 1.021 & 0.293 & 1.012 & 1.172 \\
    $R_T(A=161)$ & 1.038 & 1.005 & 3.179 & 1.036 & 0.193 \\
    $R_T(A=162)$ & 1.029 & 1.023 & 0.583 & 1.029 & 0.000 \\
    $R_T(A=163)$ & 1.020 & 1.019 & 0.098 & 1.020 & 0.000 \\
    $R_T(A=164)$ & 1.011 & 1.011 & 0.000 & 1.011 & 0.000 \\
    $R_T(A=165)$ & 1.002 & 1.002 & 0.000 & 1.002 & 0.000 \\
    $R_T(A=166)$ & 0.993 & 0.993 & 0.000 & 0.993 & 0.000 \\
    $R_T(A=167)$ & 0.984 & 0.984 & 0.000 & 0.984 & 0.000 \\
    $R_T(A=168)$ & 0.975 & 0.975 & 0.000 & 0.975 & 0.000 \\
    $R_T(A=169)$ & 0.496 & 0.496 & 0.000 & 0.496 & 0.000 \\
    $f$ & 2.286 & 2.060 & 9.886 & 2.156 & 6.684 \\
    $f_N$ & 1.000 & 1.004 & 0.400 & 1.000 & 0.000 \\
    $f_Z$ & 1.000 & 1.000 & 0.000 & 1.000 & 0.000 \\
    \end{tabular}
    \label{tab:parameters}
\end{table*}

The initial parameter covariance matrix, $\mathbb{X}$, is taken to be diagonal with a 1\% uncertainty on each parameter.  One percent is small compared to the size of the model uncertainty that we expect, but we use this small value to ensure that the linear approximation for the Kalman filter is valid. Otherwise, the full model calculations can be tens of percent off from the results of the Kalman filter.  To ensure that we allow for enough variation in the parameters, we iterate over the Kalman filter several times, where each updated $\textbf{x}_1$ becomes the $\textbf{x}_0$ for the next calculation.  In between each iteration of the Kalman filter, we run the full \beoh{} calculation, instead of using Eq. (\ref{eq:Kalman}).  This procedure allows for parameter variations greater than 1\%.  

\begin{figure*}
    \centering
    \includegraphics[width=\textwidth]{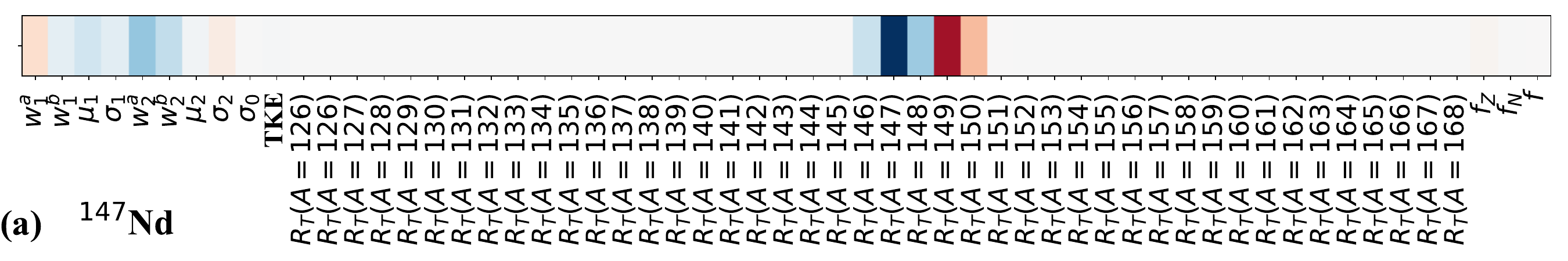} \\
    \includegraphics[width=\textwidth]{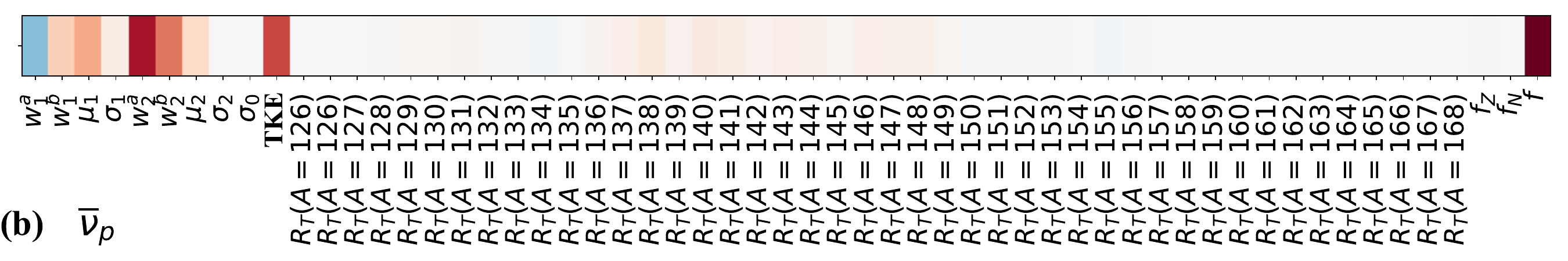}
    \caption{Sensitivities for (a) the cumulative fission product yields for $^{147}$Nd and (b) average prompt neutron multiplicity, $\overline{\nu}_p$.  Dark blue (red) regions show parameters that have large positive (negative) sensitivity.}
    \label{fig:sensitivities}
\end{figure*}

The sensitivities, Eq. (\ref{eq:pCov}), are calculated for each CFY and for $\overline{\nu}_p$.  The default parameter change that we consider is 1\%.  All parameters are increased by 1\% and decreased by 1\%, and \beoh{} is run with each of these parameter changes.  An example of the sensitivities for the CFY for $^{147}$Nd and for $\overline{\nu}_p$ is shown in Fig. \ref{fig:sensitivities} panels (a) and (b), respectively.  The regions of large positive (negative) sensitivity are shown in darker blue (red) and zero sensitivity is shown by white regions.  All of the CFYs display similar features, with a weak dependence on the pre-neutron emission mass distributions and a stronger dependence on the excitation energy sharing parameter, $R_T(A)$, for the masses near the mass of interest.  The average prompt neutron multiplicity is anti-correlated strongly with both the average total kinetic energy and the scaling factor of the spin cutoff parameter, $f$.  There are additional, weaker correlations and anti-correlations to the $Y(A)$ parameters.

\subsection{Experimental Data}


The experimental cumulative fission product yield data used in this work are taken from EXFOR \cite{EXFOR}.  Although this is a demonstration of this new procedure, to not bias the results with outlier data, we made a preliminary reading of the publications describing these data, especially when there are discrepancies between different measurements and remove inconsistent data from the fitting procedure.  We removed some or all of the data from five data sets found in EXFOR.  Two of those data sets, those of Skovorodkin \emph{et al.} \cite{Skovorodkin1973}, and Bugrov \emph{et al.} \cite{Bugrov1985} only have sparse details in the EXFOR entry, and we could not find further details of the publications; these two data sets were removed in their entirety.  For the data of Erten \emph{et al.} \cite{Erten1978}, the reported yield for $^{134}$I was given as a combination of the ground and metastable state.  Currently, we do not incorporate this type of data in our optimization procedure, so this single data point was removed from the fitting.  The data of Reber \emph{et al.} \cite{Reber2005} and Toppare \emph{et al.} \cite{Toppare1982} were removed entirely.  

The results from the Kalman filter depend strongly on the experimental uncertainties.  Uncertainties that are small will influence the fit significantly.  If these uncertainties are unrealistic, the model parameters can therefore be biased by these values, especially when the magnitude of the data is also relatively small.  For this work, we take the experimental uncertainties as reported in the EXFOR database, but drop some of the data sets during the fitting procedure due to unreasonably small uncertainties.  A more robust evaluation procedure would have to focus on identifying realistic sources and magnitudes of experimental uncertainties, such as can be done using templates of experimental uncertainties \cite{FYtemplate}.  Work by Mattera \emph{et al.} \cite{MatteraU8}, has been focusing on revising literature IFY and CFY values using current nuclear structure information and would be used in a future evaluation.

Additionally, to ensure that small CFYs would not disproportionately bias the fit, we implemented a multi-step optimization procedure.  We began by fitting the CFYs in the light and heavy peaks of the mass and charge distribution, then successively added nuclei with smaller and smaller yields.  Between each step in the optimization procedure, we performed a full \beoh{} calculation and used these CFY values as the starting point for the next optimization step.  The prior uncertainties on the parameters remain the same for each optimization step.  There were 5 steps in the optimization, where nuclei with a cumulative yield greater than 5\% were included, then 2\%, 1\%, 0.5\%, then 0.2\%.  This procedure was first performed with available experimental data, then the fit was performed to the ENDF/B-VIII.0 evaluated CFYs.  The results of these two optimizations will be compared in Sec. \ref{sec:fitComparison}.

Finally, as mentioned in Sec. \ref{sec:kalman}, the experimental covariance matrix, $\mathbb{V}$, can contain correlations between data points in addition to the experimental standard deviations.  Typically, and as far as we are aware for the CFYs used in this work, experimental covariances are not determined or reported.  Again, within the Kalman filter, these correlations can have a significant impact on the final evaluated mean values and uncertainties.  As the primary goal in this study is to describe the optimization methodology rather than performing a full evaluation, we leave it to a future work to investigate the impact that correlations would have on the Kalman filter output.  

\section{RESULTS}
\label{sec:results}

In this section, we discuss the results obtained in two different approaches:  first by fitting to experimental data in Sec. \ref{sec:fitExperiment} and to ENDF/B-VIII.0 in Sec. \ref{sec:fitENDF}. We then compare the results of the two optimizations in Sec. \ref{sec:fitComparison}. Finally, we discuss the consistency of our FPY results with other prompt and delayed observables in Sec. \ref{sec:observables}.

\subsection{Fit to Experimental Data}
\label{sec:fitExperiment}

We first fit directly to available data, including CFYs with yields of at least 0.2\%, and $\overline{\nu}_p$.  The average prompt neutron multiplicity is included in the optimization to constrain the average total kinetic energy which otherwise moves to an unphysical value during the optimization.  The final fitted parameters are shown in the third column of Table \ref{tab:parameters}.  The parameters of the multi-Gaussian distribution $Y(A)$ change by less than 10\%, although the largest changes come in the width of the first Gaussian, $\sigma_1$, and the mean of the second Gaussian, $\mu_2$.  Additionally, we see no change in the $R_T$ values for the highest and lowest masses, e.g. $A \leq 128$ and $A\geq 164$, because the CFYs that would be sensitive to these values were not included in the optimization, due to the imposed 0.2\% cutoff.  It was also shown in \cite{Okumura2022} that $\overline{\nu}_d$ shows significant sensitivity to the Wahl scaling factors, $f_N$ and $f_Z$, which amplify the odd-even staggering in the initial post-scission fragment yields.  However, we did not include $\overline{\nu}_d$ in the current fit.

\begin{figure}
    \centering
    \includegraphics[width=\columnwidth]{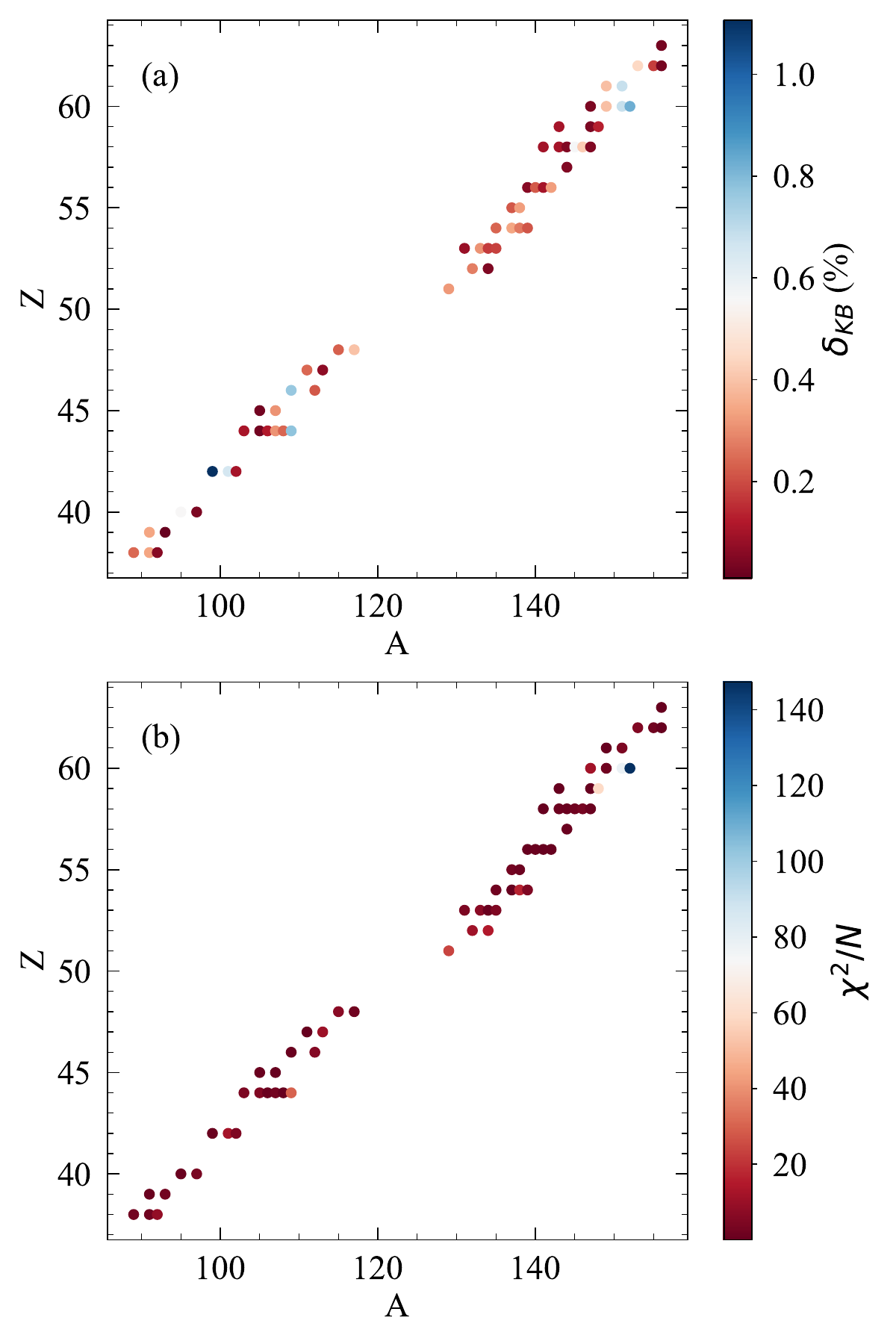}
    \caption{(a) Percent difference between the Kalman filter results and \beoh{} in the last optimization step for the fit to experimental data, as defined in Eq. (\ref{eqn:KalmanComp}).  (b) Goodness of fit, defined by $\chi^2/N$, for the fit to experimental data.}
    \label{fig:fitQualityData}
\end{figure}

The total $\chi^2/N$ for this optimization was 6.158 and the comparison between \beoh{} and the Kalman filter FPYs in this last step gave $\delta_{KB}=0.27\%$ from Eq. (\ref{eqn:KalmanComp}).  If the percent difference between the Kalman filter and \beoh{}-calculated CFYs had been on the order of a few percent, we would have run additional Kalman filter calculations until the agreement between these outputs and the full \beoh{} calculation was within a few percent.  We do this to ensure that we will not introduce erroneous values into the final evaluated values coming from the model calculation.  Figure \ref{fig:fitQualityData}(a) shows the percent difference, Eq. (\ref{eqn:KalmanComp}), between the Kalman filter and the full \beoh{} calculations.  Almost all comparisons are better than 1\%.  However, when pushing the FPY yield limit to values lower than 0.2\%, we saw a significant uptick in $\delta_{KB}$.  The largest differences came from $Z=50,51$, tin and antimony, but removing these elements from the fitting did not lower the overall $\delta_{KB}$.  

\begin{figure*}
    \centering
    \begin{tabular}{ccc}
    \includegraphics[width=0.33\textwidth]{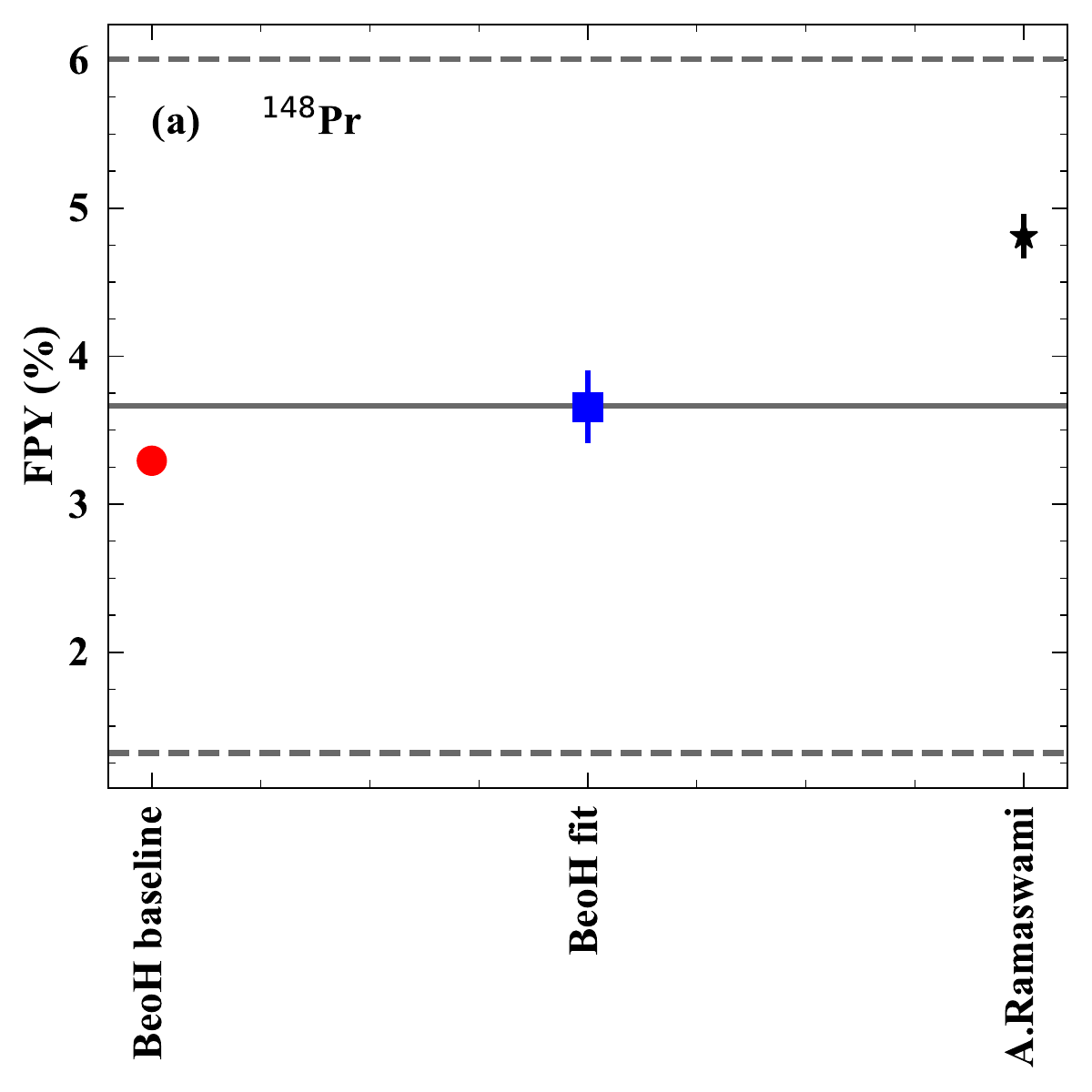} & \includegraphics[width=0.33\textwidth]{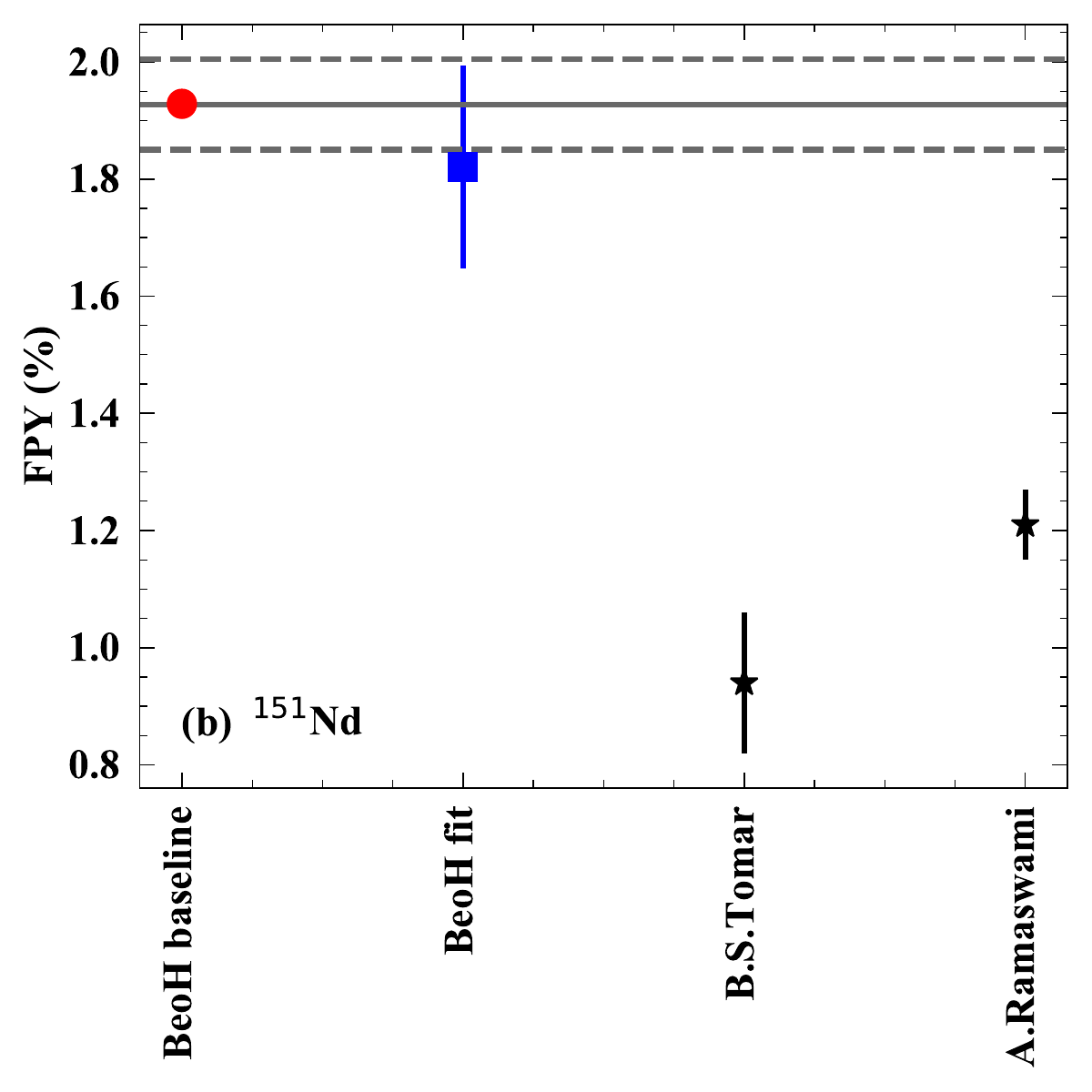} & \includegraphics[width=0.33\textwidth]{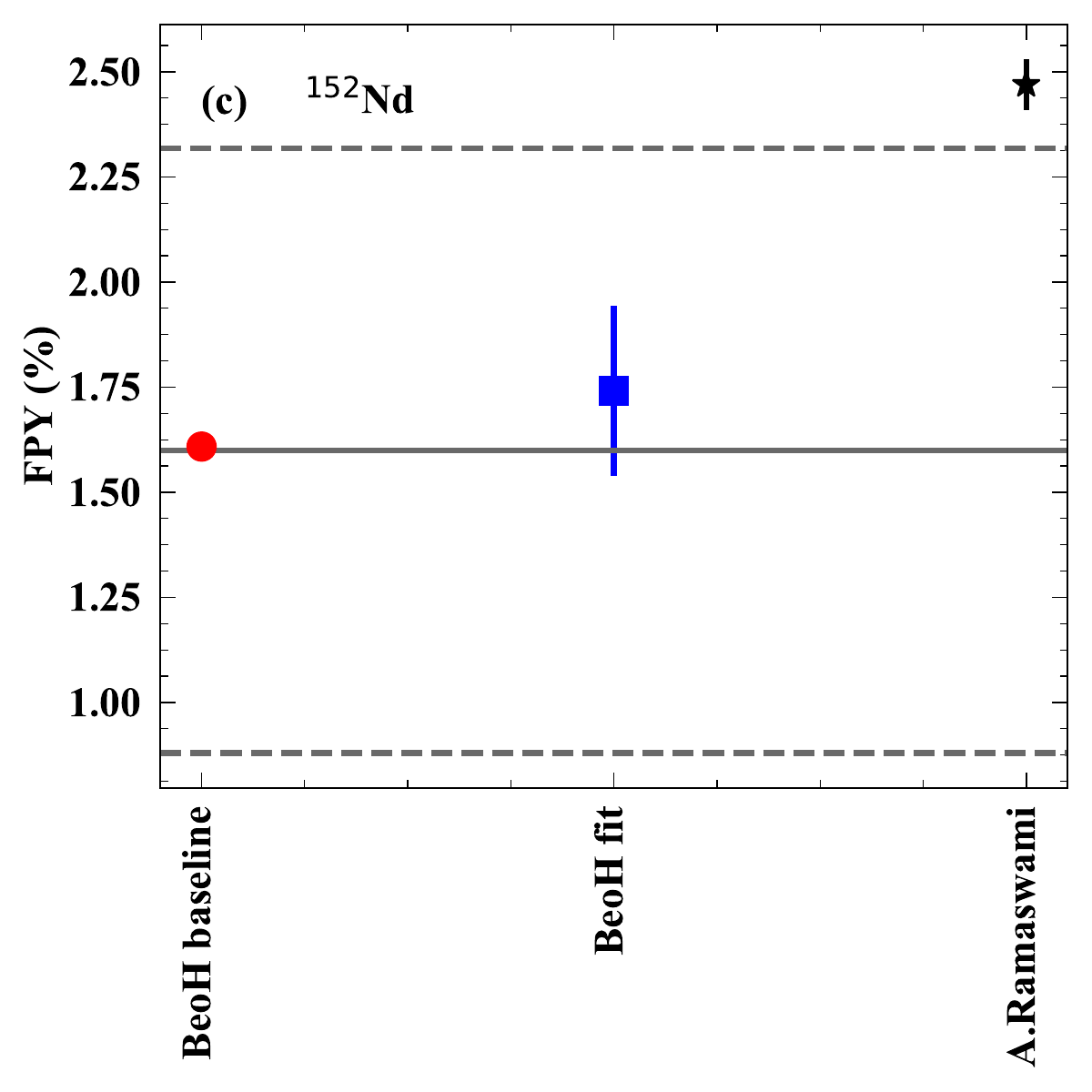}
    \end{tabular}
    \caption{Comparison between the cumulative fission product yields of (a) $^{148}$Pr, (b) $^{151}$Nd, and (c) $^{152}$Nd, for the initial \beoh{} calculation (red circle), the fit to experimental data (blue square), and experimental data (black stars, first author given on the x axis).  The gray solid line gives the mean value from the ENDF/B-VIII.0 evaluation with $1\sigma$ uncertainties shown by the gray dashed lines.}
    \label{fig:highChi2}
\end{figure*}

\begin{figure*}
    \centering
    \begin{tabular}{ccc}
    \includegraphics[width=0.33\textwidth]{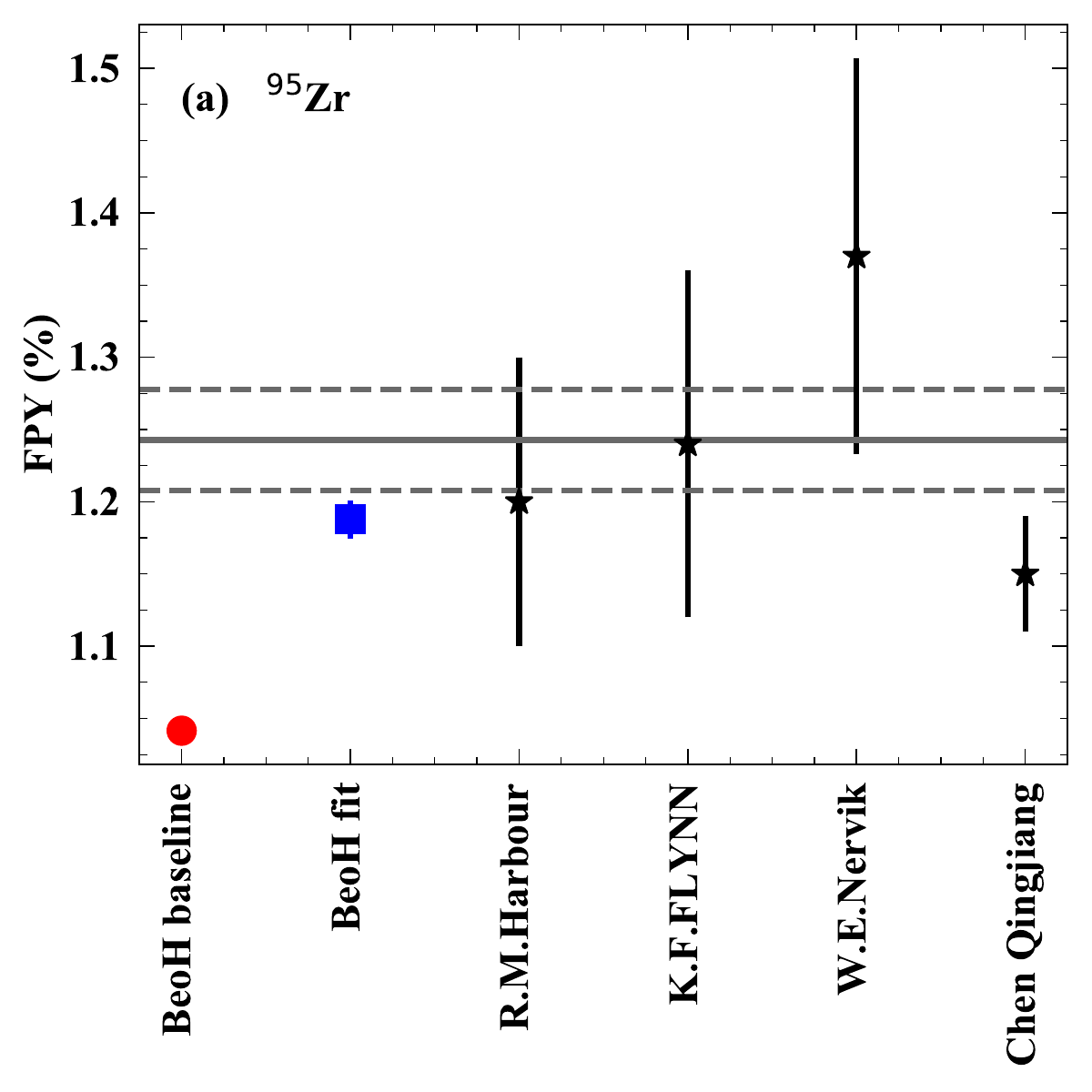} & \includegraphics[width=0.33\textwidth]{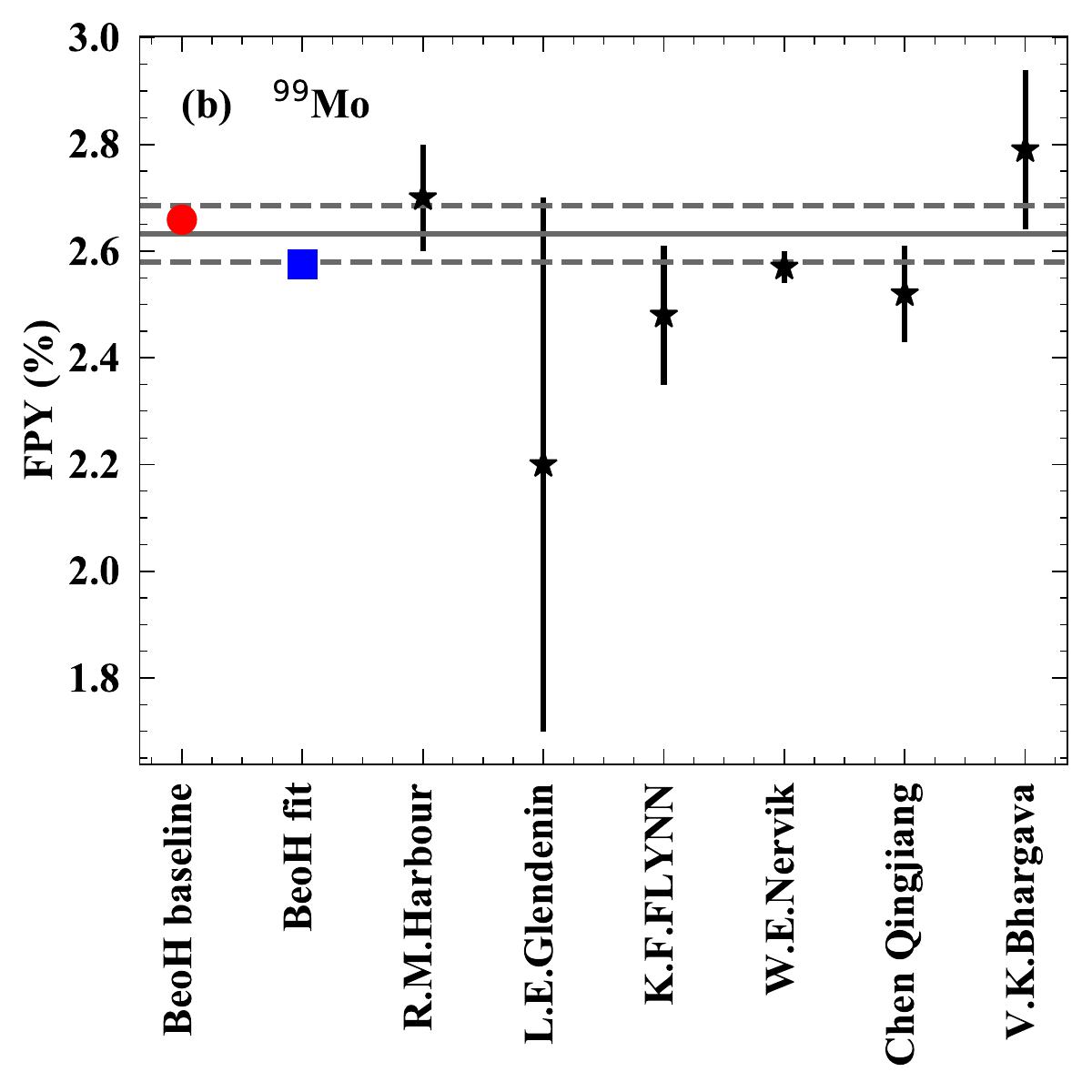} & \includegraphics[width=0.33\textwidth]{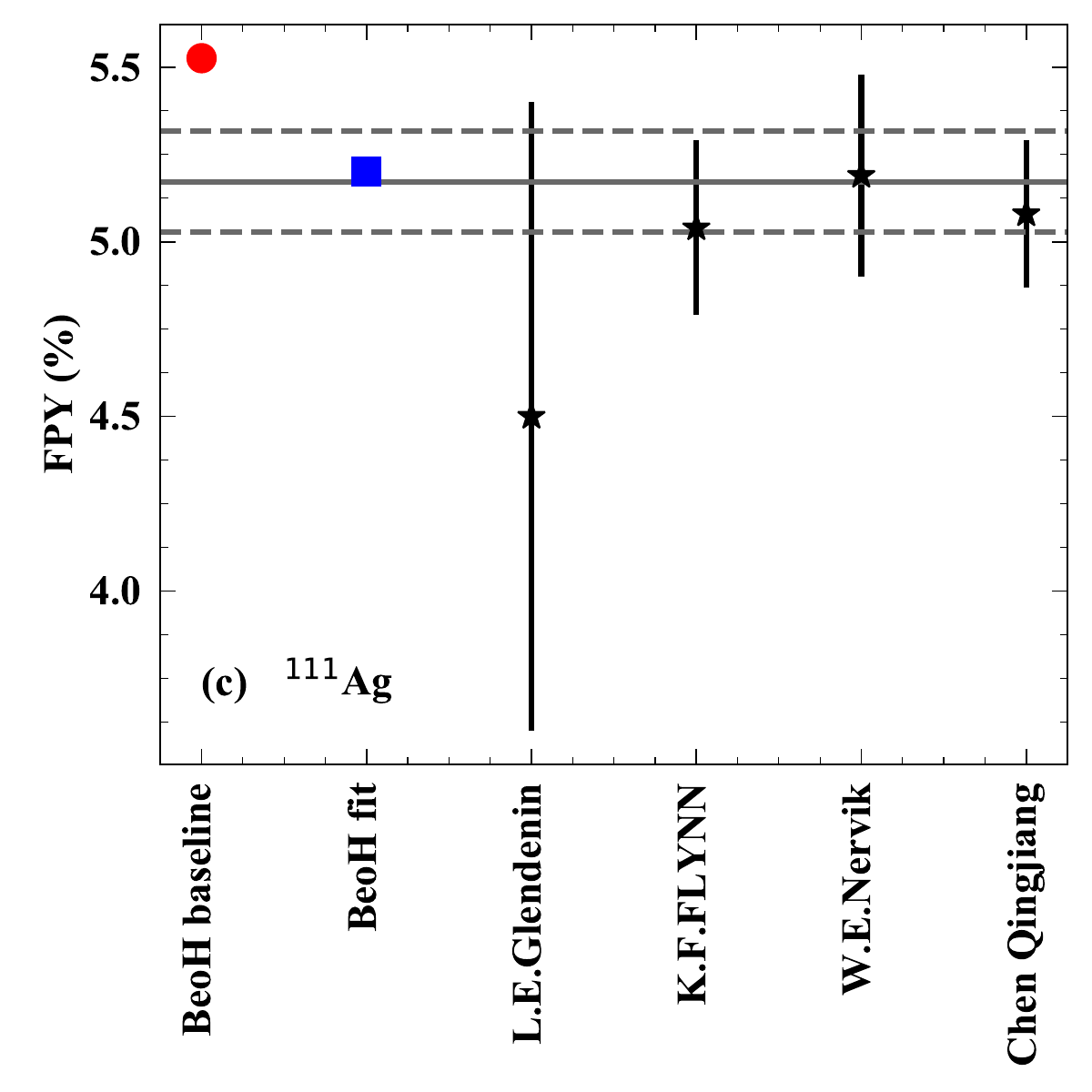} \\
    \includegraphics[width=0.33\textwidth]{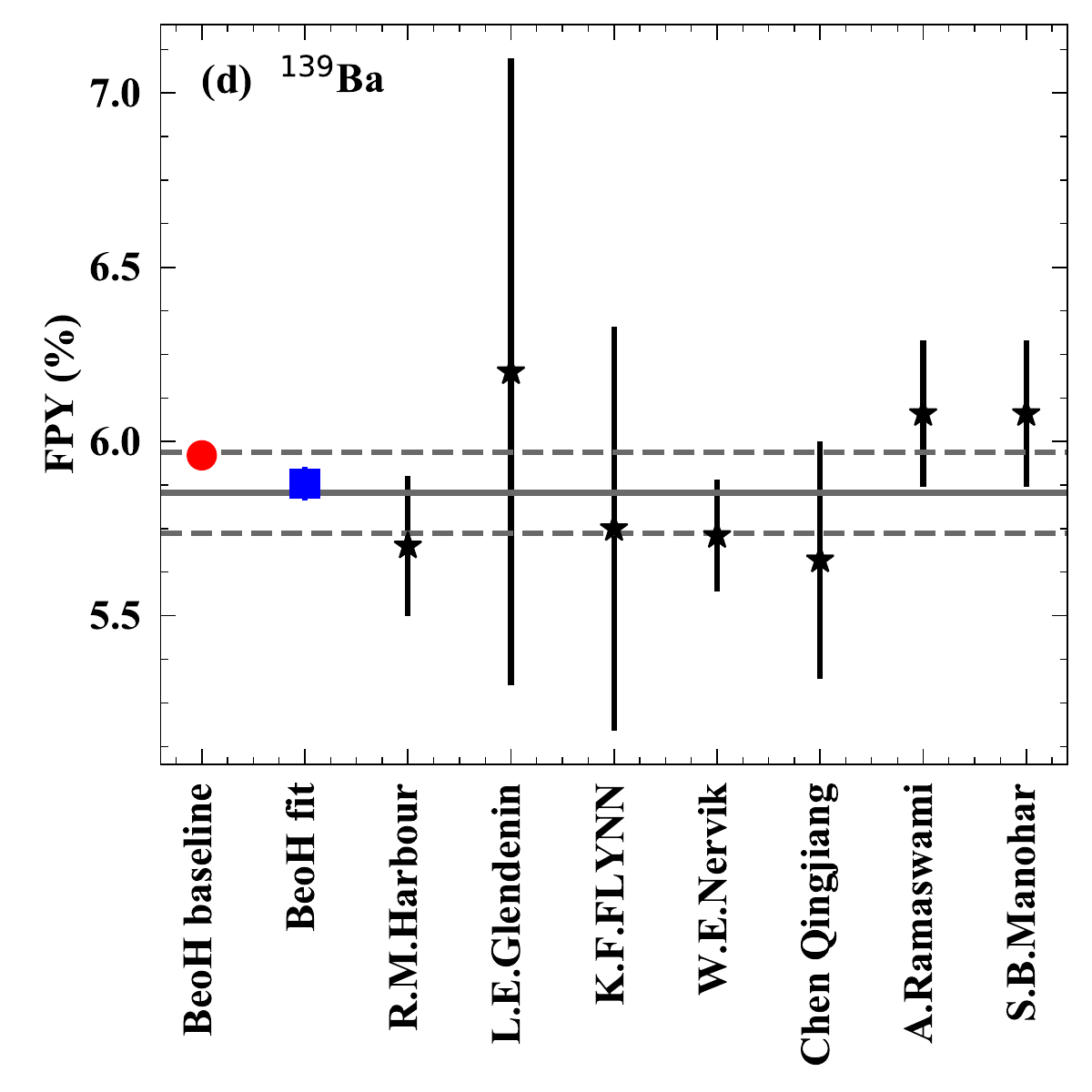} & \includegraphics[width=0.33\textwidth]{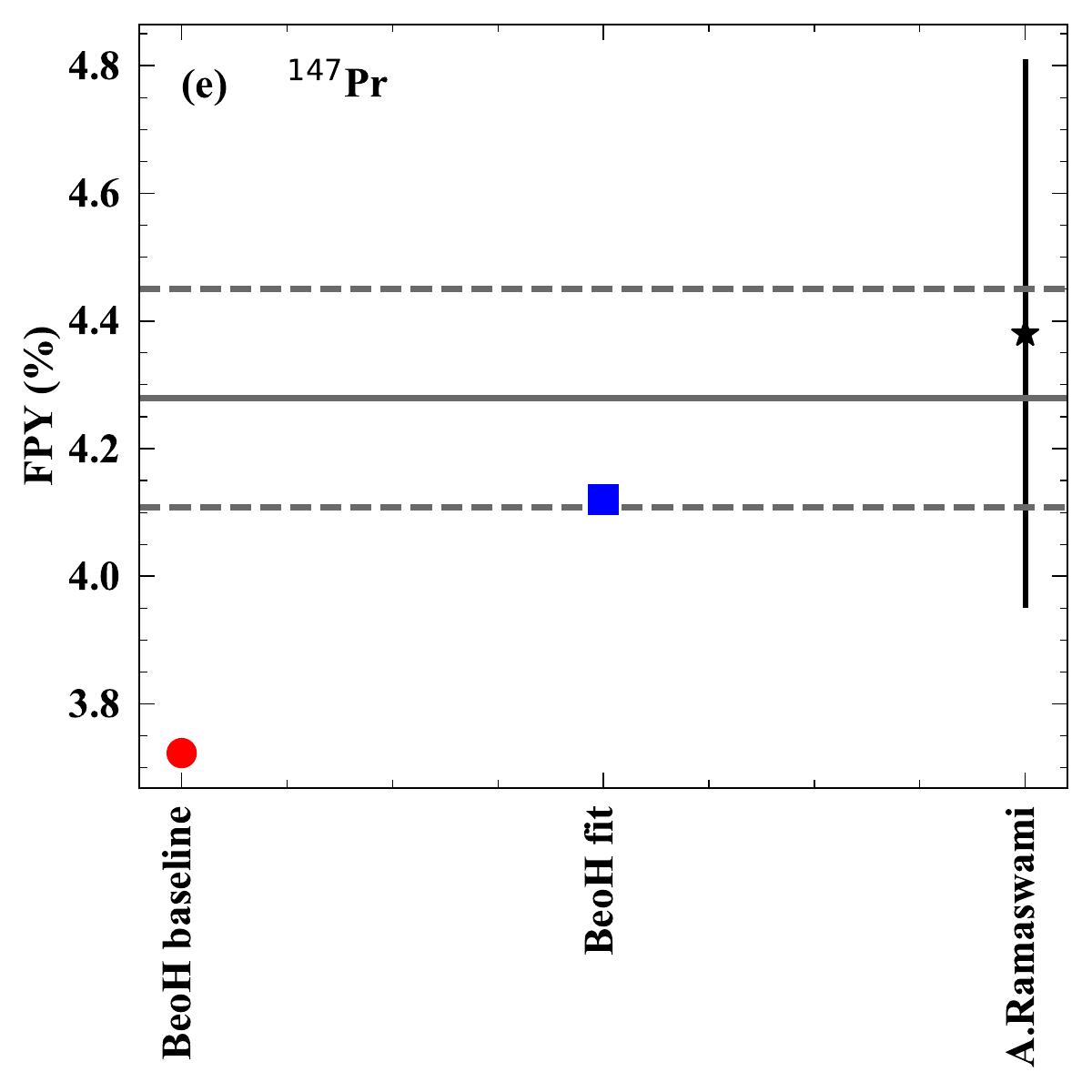} & \includegraphics[width=0.33\textwidth]{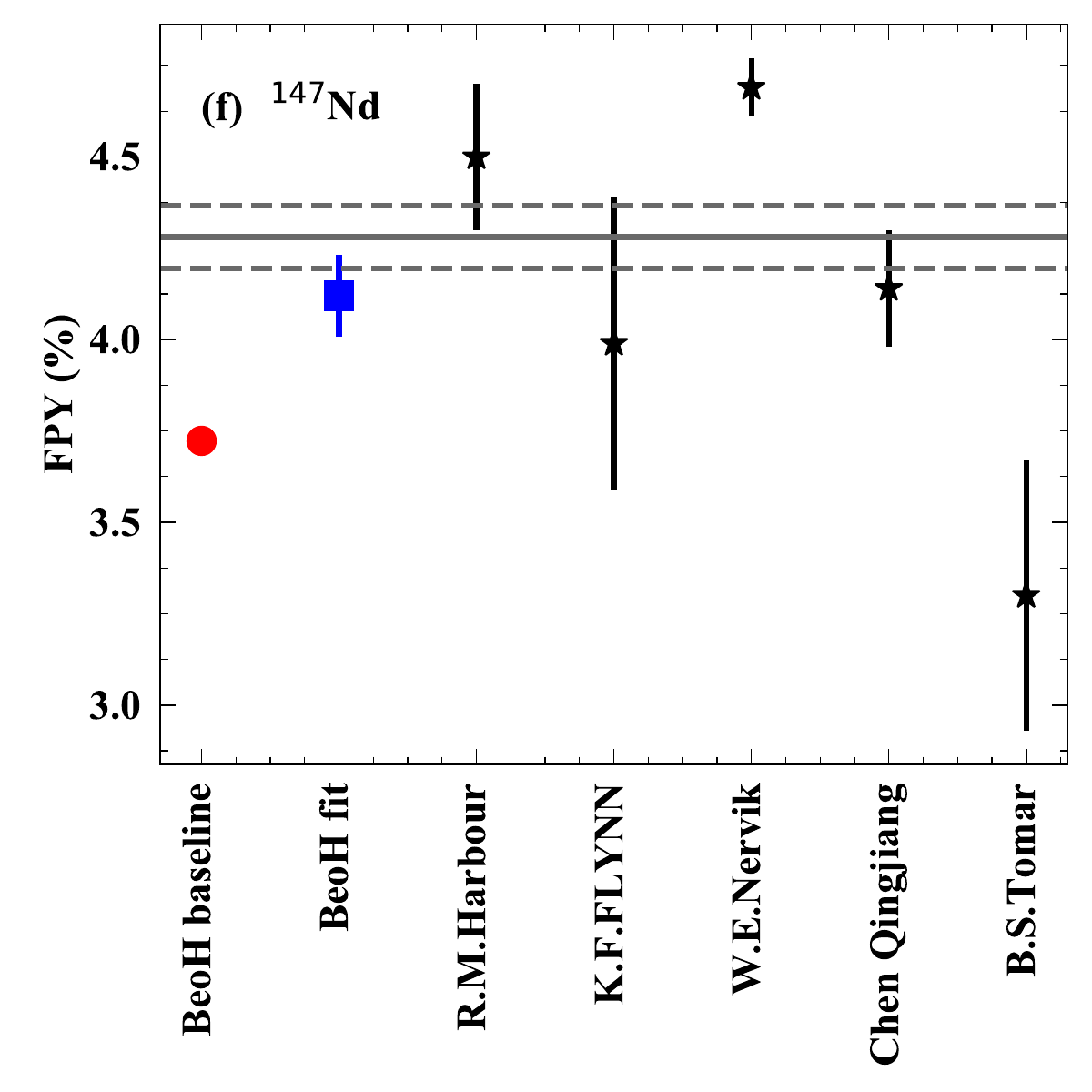}
    \end{tabular}
    \caption{Same as Fig. \ref{fig:highChi2} for (a) $^{95}$Zr, (b) $^{99}$Mo, (c) $^{111}$Ag, (d) $^{139}$Ba, (e) $^{147}$Pr, and (f) $^{147}$Nd.}
    \label{fig:dataFPY}
\end{figure*}

In Fig. \ref{fig:fitQualityData}(b), we show the $\chi^2/N$ for each of the nuclei included in the optimization.  The largest $\chi^2/N$ values are for $^{148}$Pr and $^{151,152}$Nd.  The comparisons among the initial and fitted \beoh{} calculations (red circles and blue squares, respectively), ENDF/B-VIII.0 evaluation (horizontal solid gray lines with uncertainties given by the dashed lines), and experimental data for these three nuclei are shown in Fig. \ref{fig:highChi2}.  For $^{148}$Pr, the data of Ramaswami \emph{et al.} is within the large uncertainties of the ENDF/B-VIII.0 evaluation; however, the \beoh{} calculation is about 20\% lower than the data.  For the two Nd CFYs, the data is outside of the evaluated one-sigma uncertainties.  For these three nuclei, the original \beoh{} calculations are within the evalauted uncertainties, and the fitted \beoh{} calculations are either closer to the ENDF/B-VIII.0 mean values or still within uncertainties, even though ENDF/B-VIII.0 was not included in the optimization.  In Fig. \ref{fig:dataFPY}, we show several examples of the CFYs that better reproduce the experimental data.  There is overall good agreement between the optimized \beoh{} calculation and the spread of the experimental data, as additionally indicated by the $\chi^2/N$ values in Fig. \ref{fig:fitQualityData}(b).  

\subsection{Fit to ENDF/B-VIII.0}
\label{sec:fitENDF}

\begin{figure}
    \centering
    \includegraphics[width=\columnwidth]{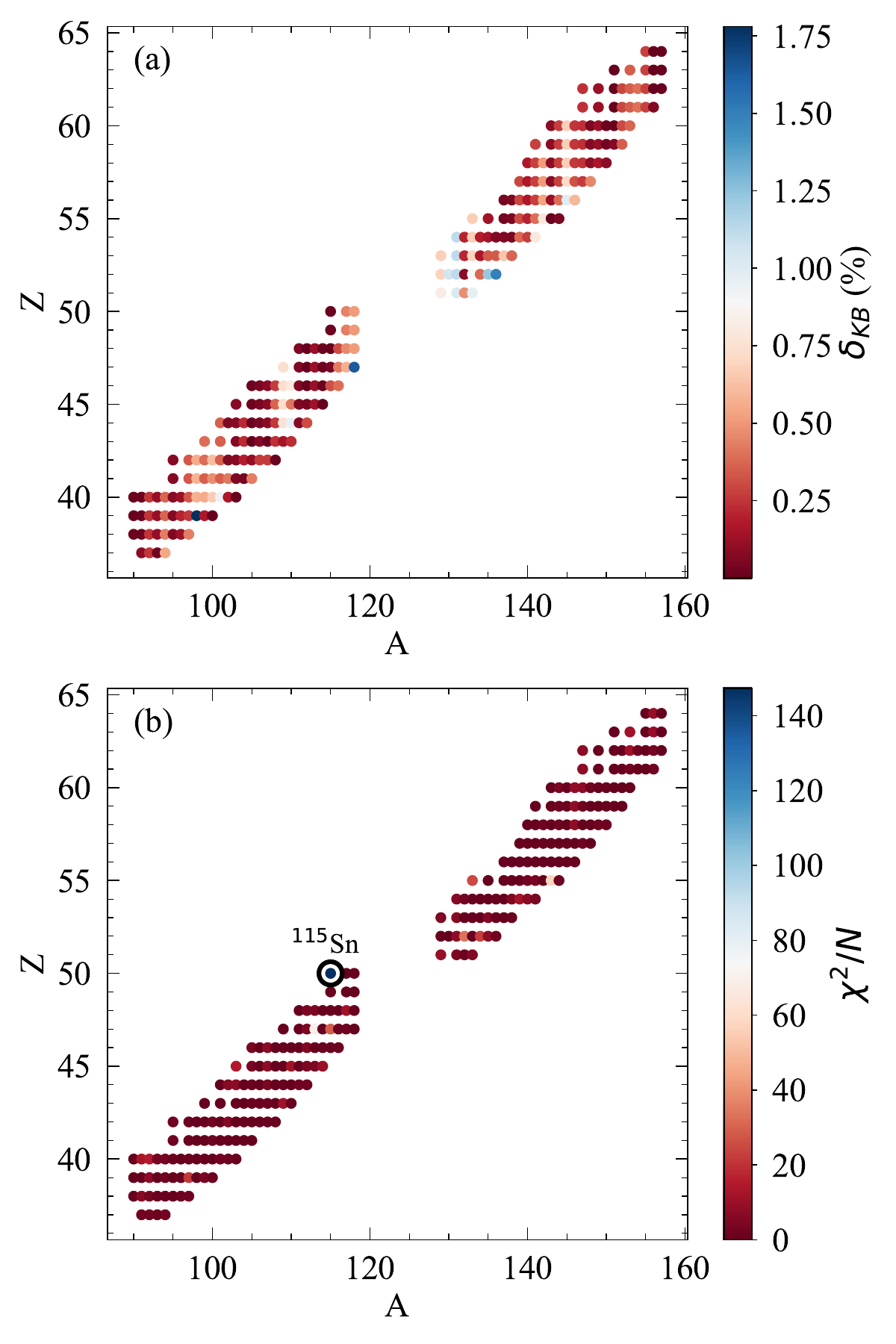}
    \caption{Same as Fig. \ref{fig:fitQualityData} for the fit to ENDF/B-VIII.0.  In (b), we put a cutoff on the maximum $\chi^2/N$ value from Fig. \ref{fig:fitQualityData}(b); the only $\chi^2/N$ value higher than this is $^{115}$Sn.  See text for details.}
    \label{fig:fitQualityENDF}
\end{figure}

Alternatively, we fit the \beoh{}-calculated CFYs to the evaluated values from ENDF/B-VIII.0.  We begin again with the baseline calculation and perform an iterative optimization, first including CFYs with yields of at least 5\%, then 1\%, then 0.5\%.  We also remove the ENDF values for $^{99}$Ru and $^{113}$In from the fitting procedure because the ENDF/B-VIII.0 values are much larger than the values from \beoh{}, which can adversely impact the optimization.  Because we did not find any experimental data for these two CFYs, it is difficult to say which is more likely to be true.  Although when we fit the experimental data we could push the yields down to 0.2\% without ruining the agreement between the Kalman filter and \beoh{}, $\delta_{KB}$ would jump from 0.30\% to 7.15\%, with the largest differences coming from the peaks of the mass distribution around masses 103 and 147.  These differences indicate that, again, the smallest CFYs are having the strongest influence on the optimization, destroying the agreement with the largest CFYs, which should be the easiest to reproduce with the most reliable data.  Additionally, we have not taken the parameter correlations from the previous iteration as the starting correlations for the next iteration, which could potentially also improve the stepped fitting approach and should be investigated in the future.  In Fig. \ref{fig:fitQualityENDF}, we show (a) $\delta_{KB}$ and (b) $\chi^2/N$ for the fit to ENDF/B-VIII.0 for cumulative CFYs with yields at least 0.5\%.  The overall agreement between \beoh{} and the Kalman filter for the last step in the optimization is very good, with $\delta_{KB} = 0.30\%$, similar to the fit with experimental data in Sec. \ref{sec:fitExperiment}.

\begin{figure*}
    \centering
    \begin{tabular}{ccc}
    \includegraphics[width=0.33\textwidth]{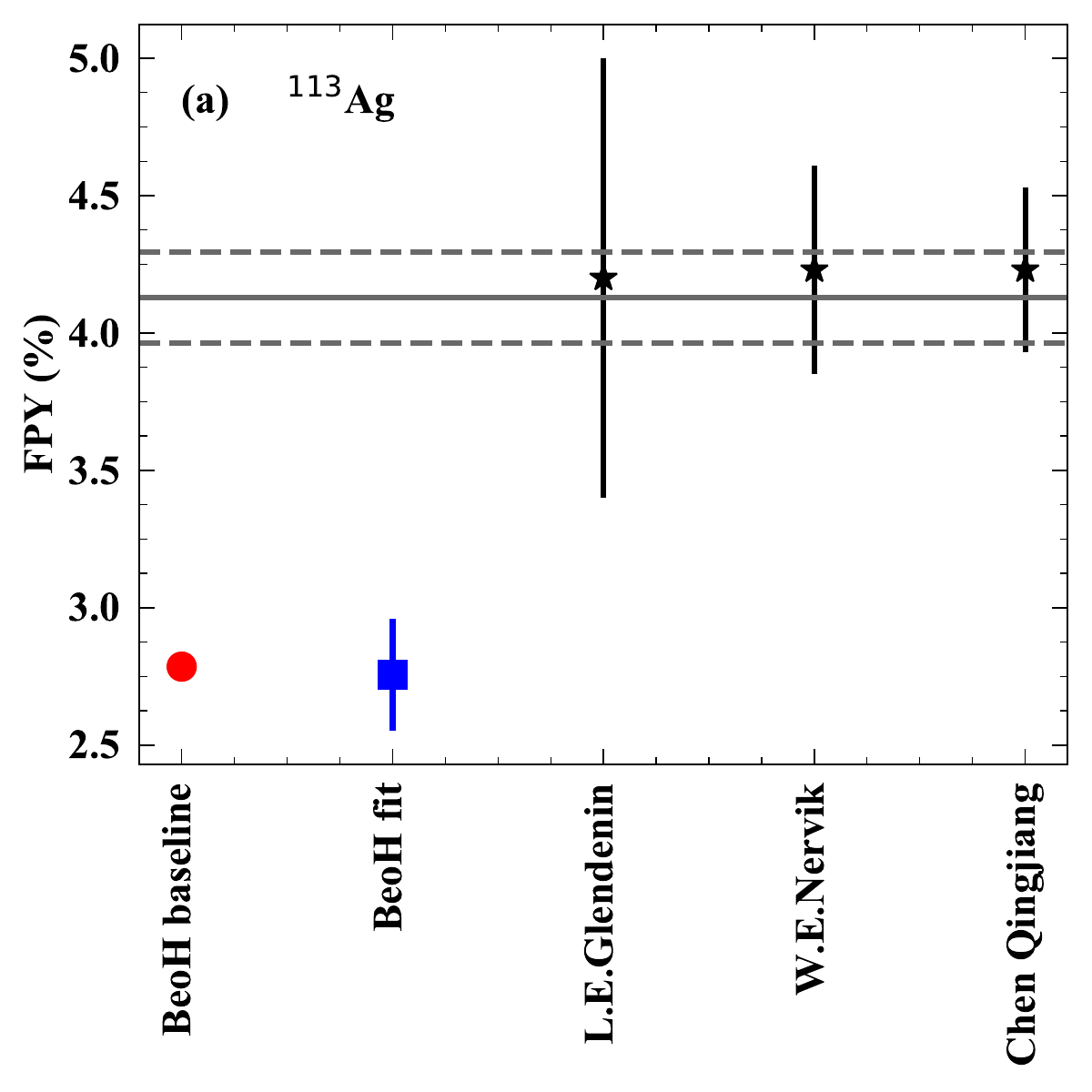} & \includegraphics[width=0.33\textwidth]{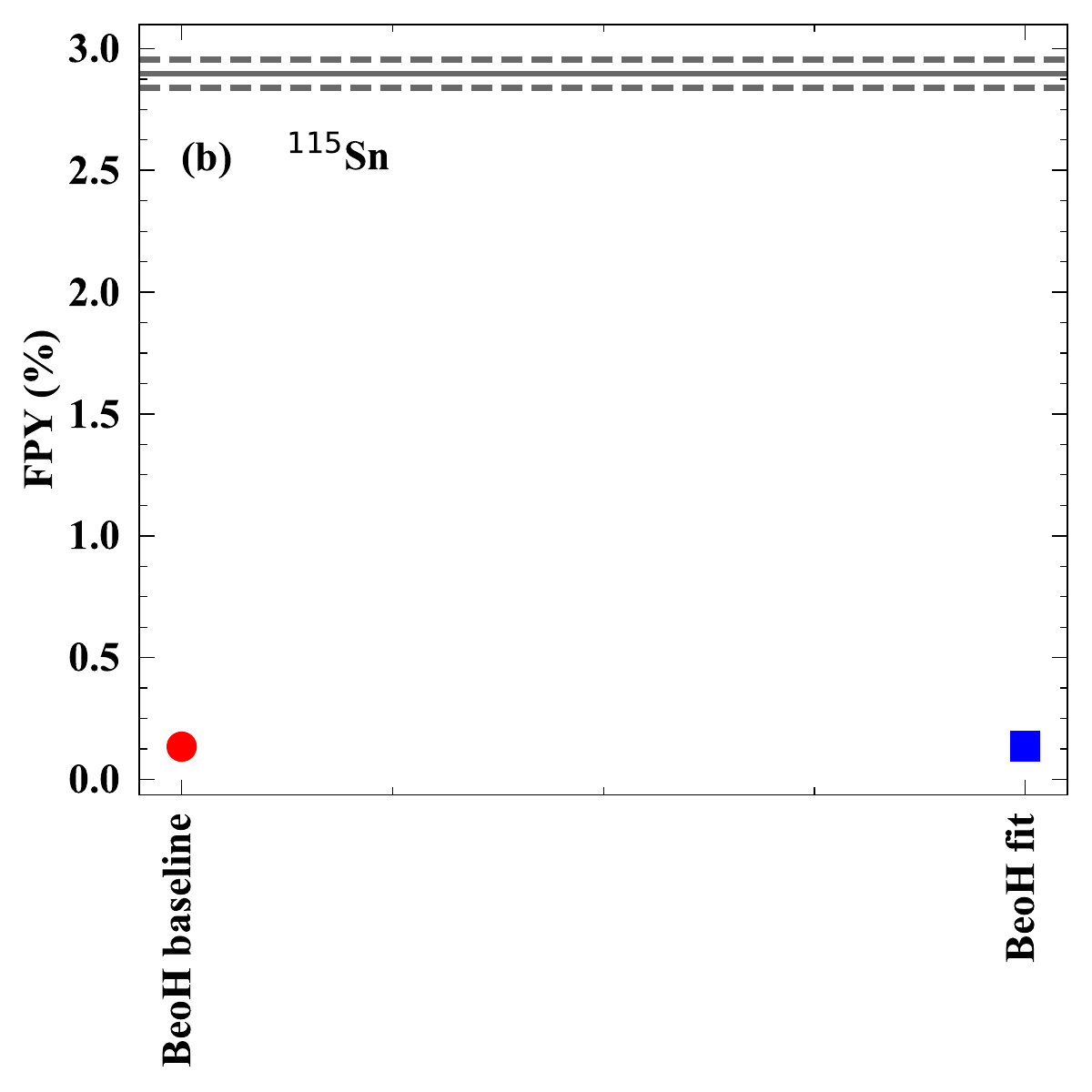} & \includegraphics[width=0.33\textwidth]{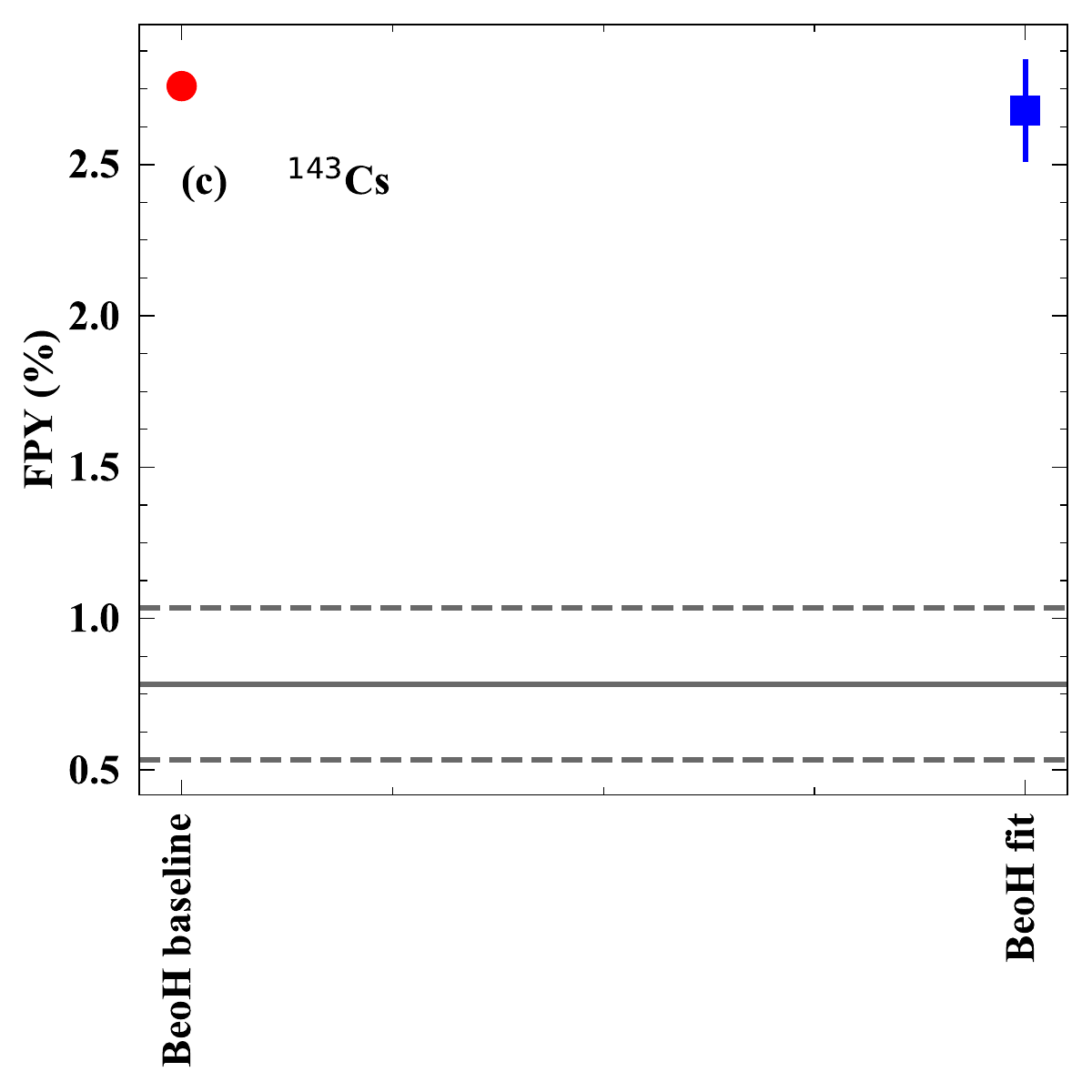} 
    \end{tabular}
    \caption{Same as Fig. \ref{fig:highChi2} where the blue squares give the fit the ENDF/B-VIII.0 for (a) $^{113}$Ag, (b) $^{115}$Sn, and (c) $^{143}$Cs.}
    \label{fig:highChi2ENDF}
\end{figure*}

The $\chi^2/N$ values are all below 70, except for $^{115}$Sn where $\chi^2/N=2271$.  The next two highest $\chi^2/N$ values are for $^{113}$Ag and $^{143}$Cs, almost two orders of magnitude lower.  The comparisons among the initial \beoh{} calculation, the fitted \beoh{} calculation, ENDF/B-VIII.0, and experimental data are shown in Fig. \ref{fig:highChi2ENDF}.  Clearly, the large $\chi^2/N$ for $^{115}$Sn--panel (b)--is because the \beoh{} calculation is about an order of magnitude smaller than ENDF/B-VIII.0.  There is no experimental data available for this CFY.  For $^{113}$Ag and $^{143}$Cs, the calculated \beoh{} CFY is not quite as discrepant from the ENDF/B-VIII.0 values as for $^{115}$Sn.  There are data available for $^{113}$Ag, shown in Fig. \ref{fig:highChi2ENDF}(a) (black stars), which are consistent with the evaluated value.  

\begin{figure*}
    \centering
    \begin{tabular}{ccc}
    \includegraphics[width=0.33\textwidth]{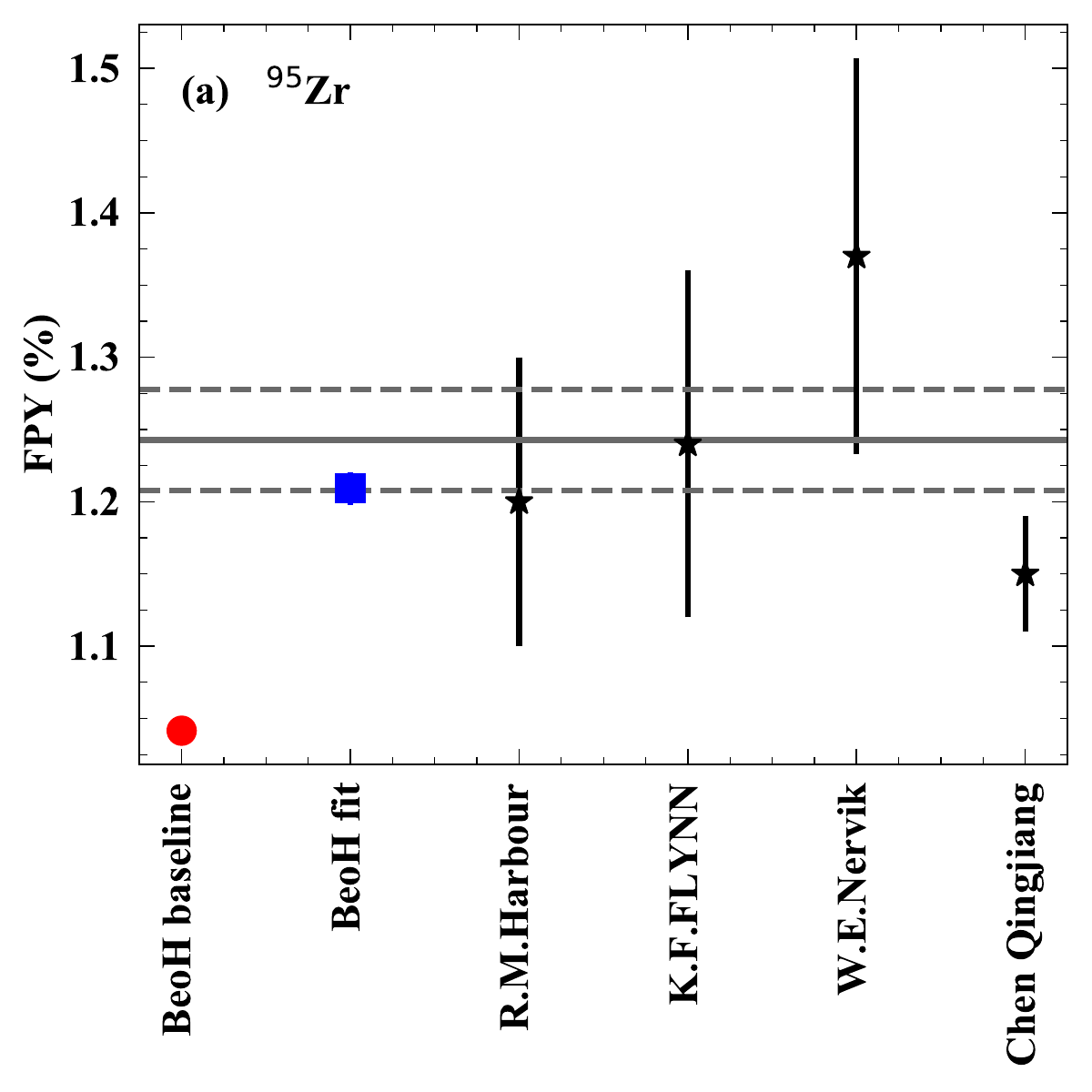} & \includegraphics[width=0.33\textwidth]{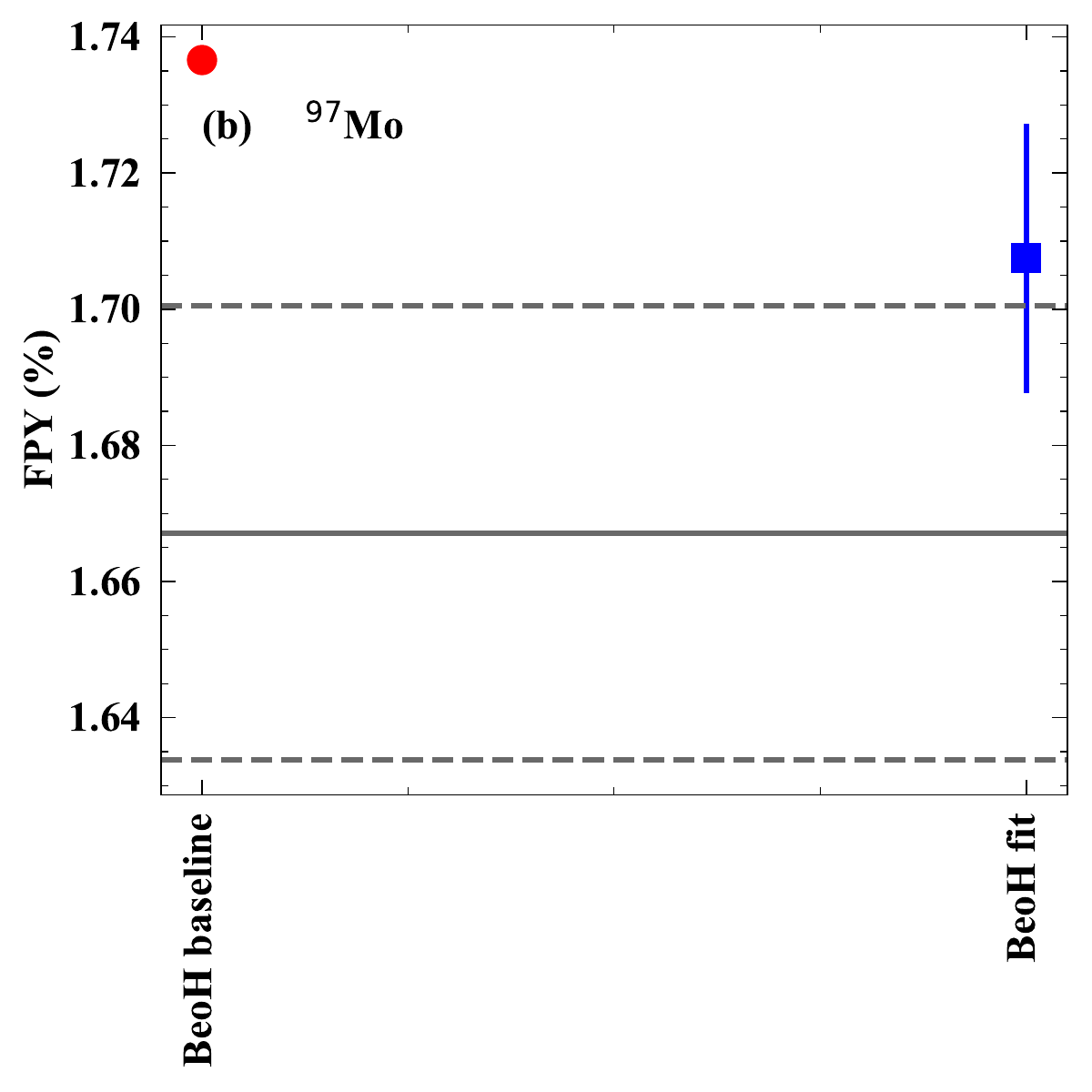} & \includegraphics[width=0.33\textwidth]{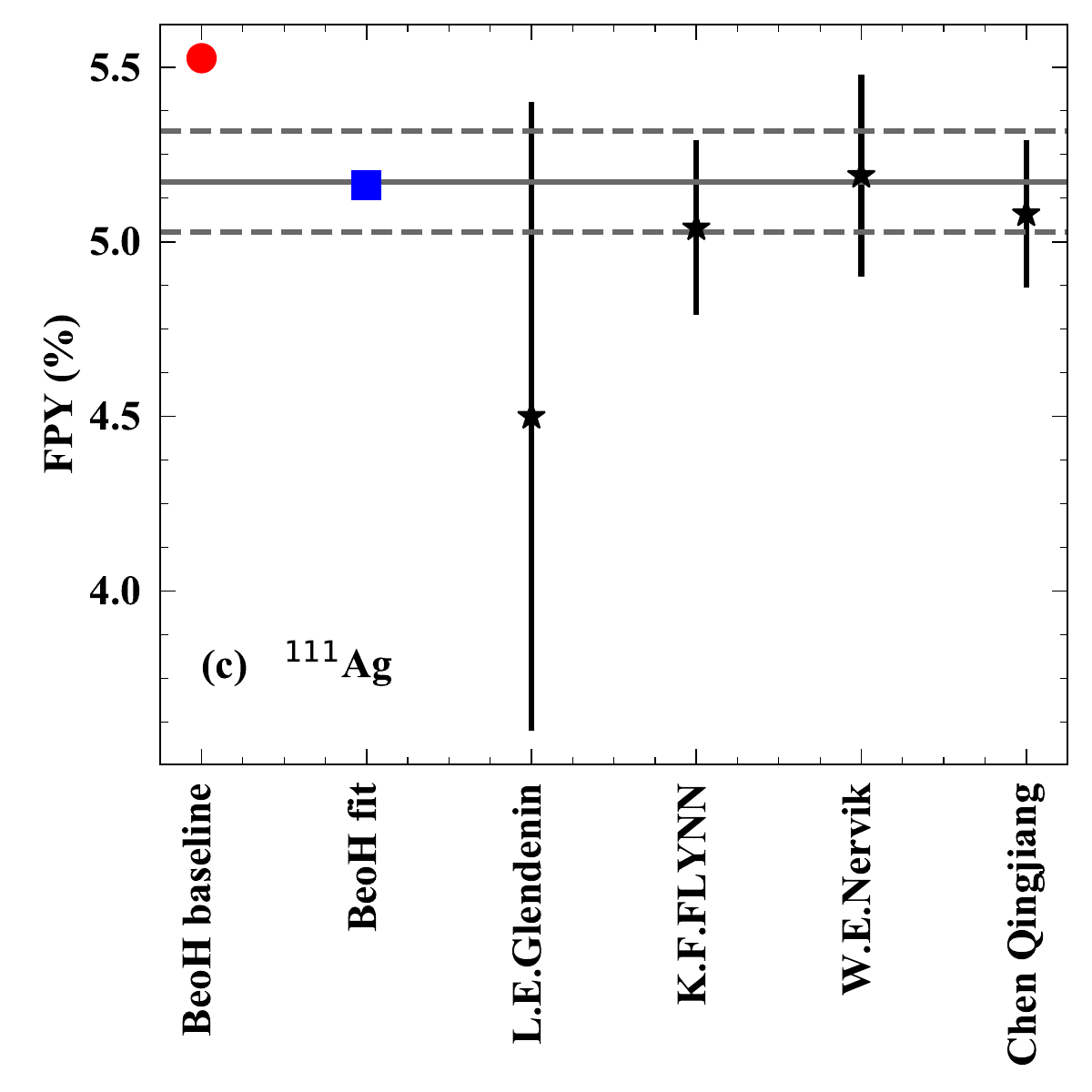} \\ 
    \includegraphics[width=0.33\textwidth]{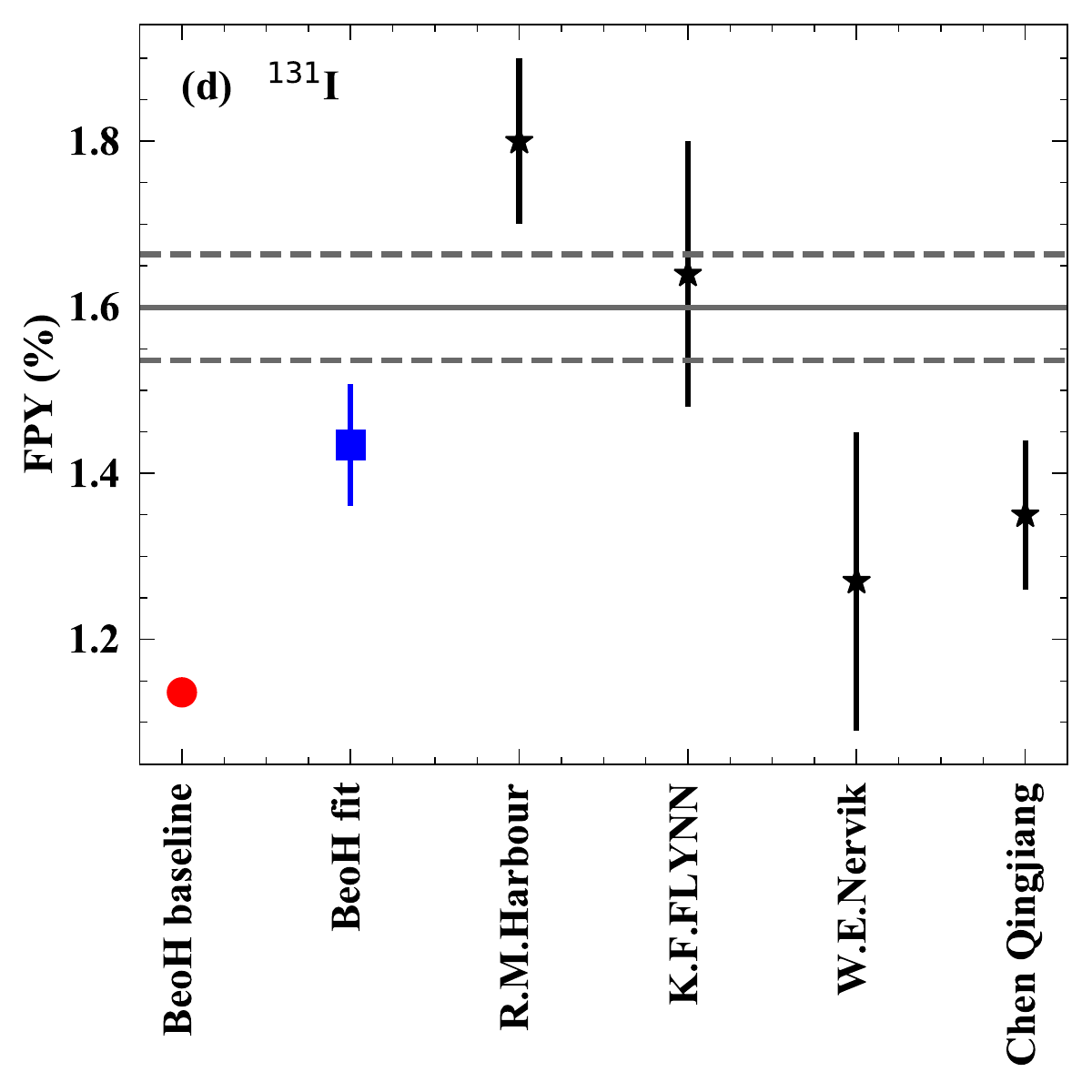} & \includegraphics[width=0.33\textwidth]{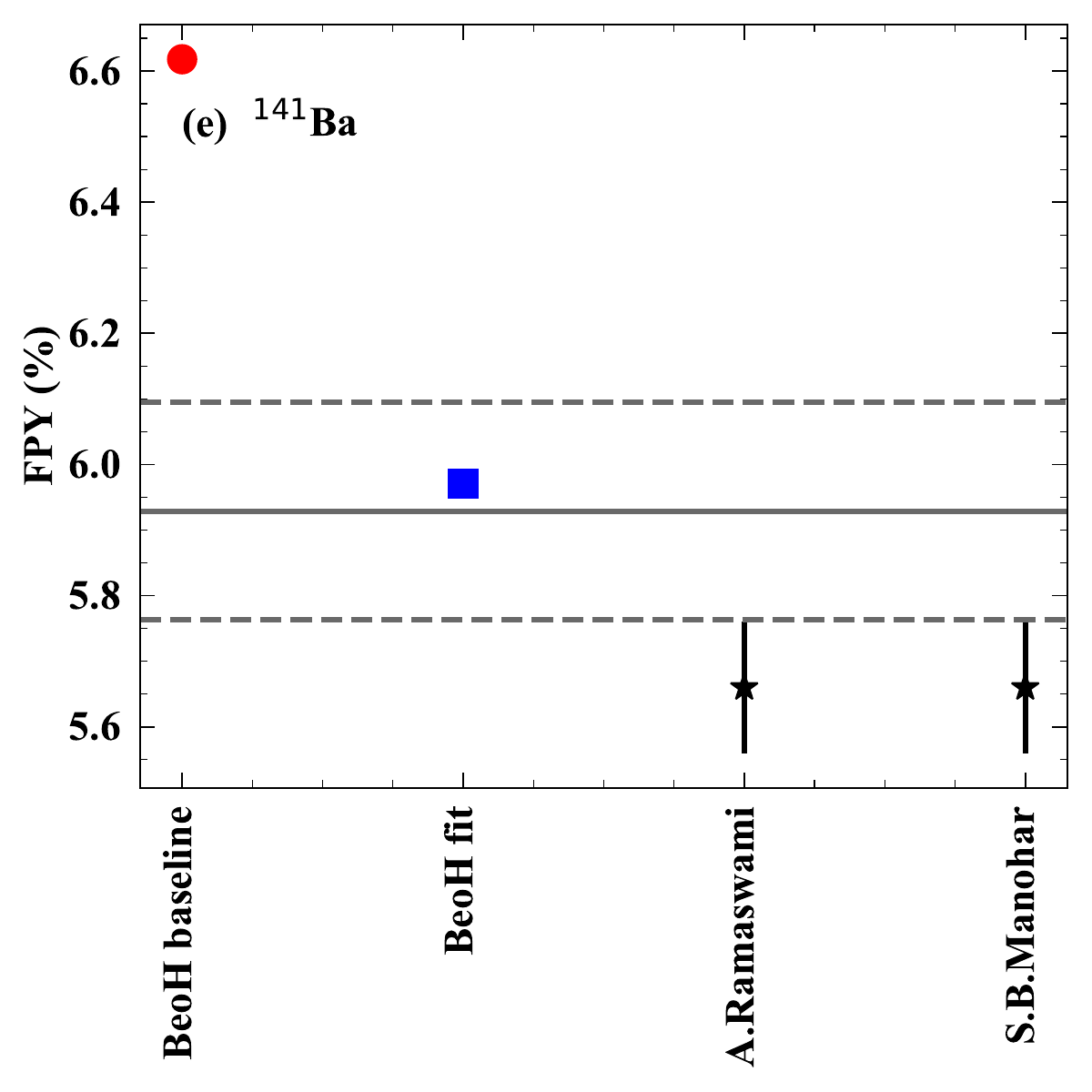} & \includegraphics[width=0.33\textwidth]{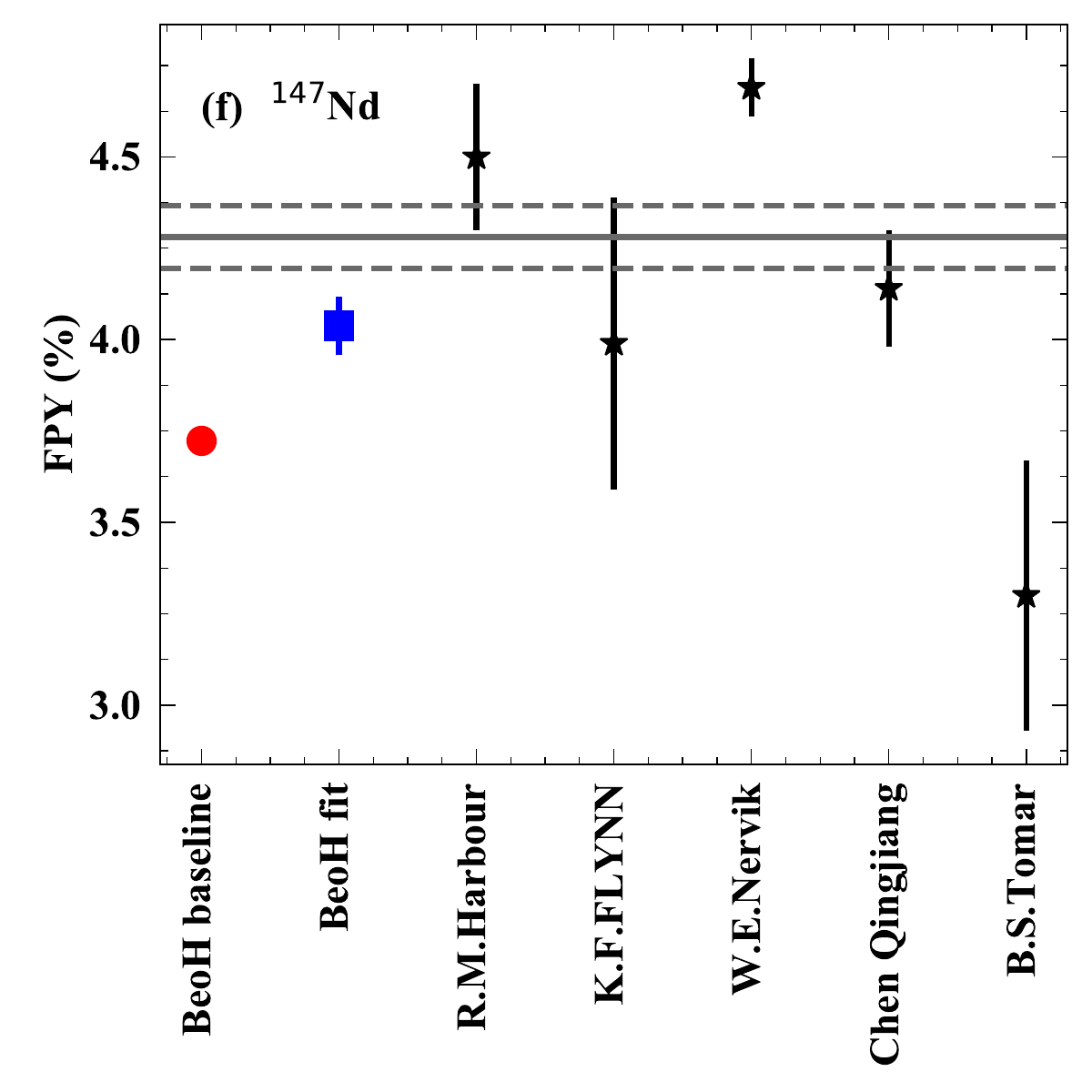} 
    \end{tabular}
    \caption{Same as Fig. \ref{fig:highChi2ENDF} (a) $^{95}$Zr, (b) $^{97}$Mo, (c) $^{111}$Ag, (d) $^{131}$I, (e) $^{141}$Ba, and (f) $^{147}$Nd.}
    \label{fig:dataFPYENDF}
\end{figure*}

Then, in Fig. \ref{fig:dataFPYENDF}, we show a handful of other examples of the CFYs calculated from \beoh{} before and after the optimization procedure.  These values are compared both to the ENDF/B-VIII.0 evaluation and available experimental data.  The Kalman filter optimization was again able to adjust the \beoh{} calculations towards the ENDF/B-VIII.0 CFY values.  

\subsection{Comparison Between Optimizations}
\label{sec:fitComparison}

\begin{figure}
    \centering
    \includegraphics[width=\columnwidth]{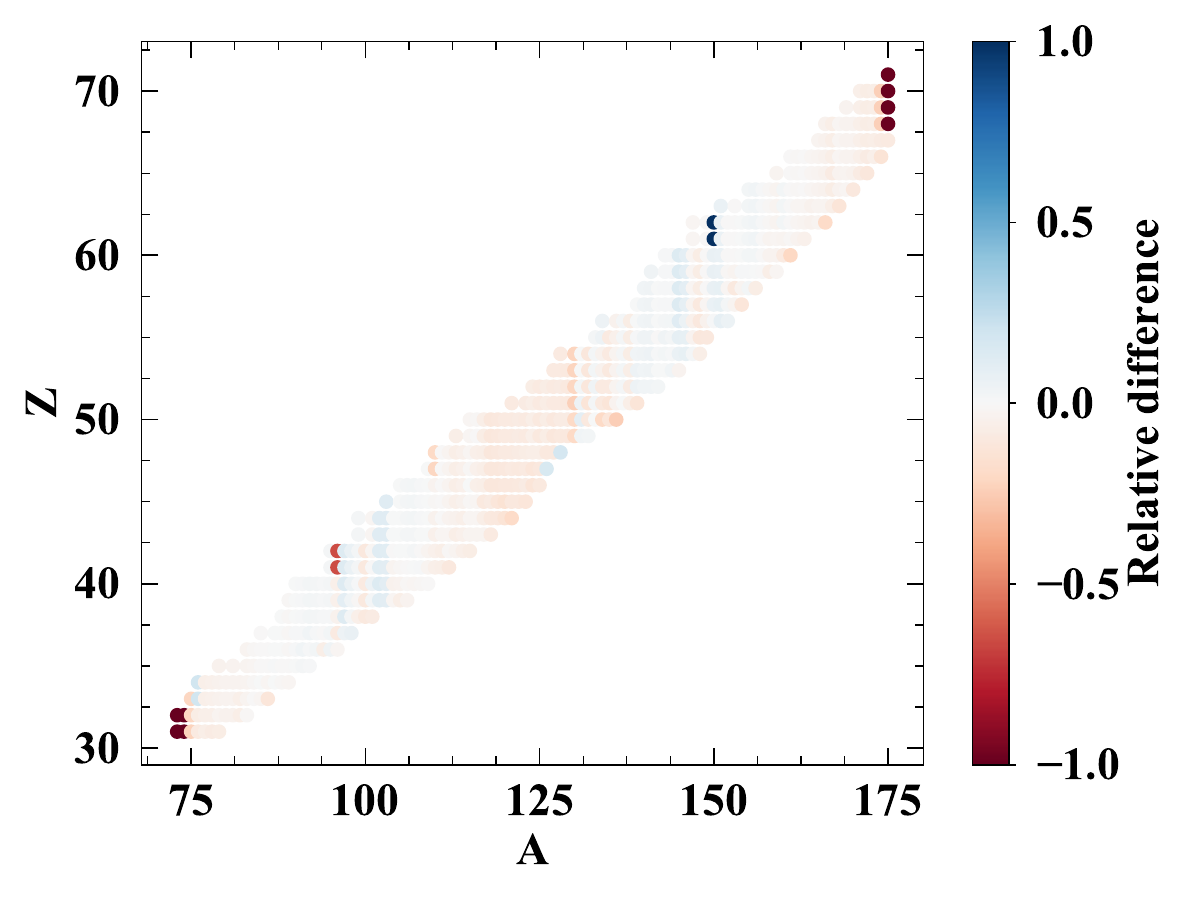}
    \caption{Relative difference, as defined in Eq. (\ref{eqn:relDiff}), between the calculated CFYs fitted to EXFOR and ENDF/B-VIII.0 data.}
    \label{fig:fitComp}
\end{figure}

We can now directly compare the two optimizations.  We expect that differences come from two main sources:  the different CFYs that are included in the two optimizations and that the experimental data can be inconsistent with the evaluated values.  In Fig. \ref{fig:fitComp}, we show an overview of the CFYs from both final calculations by plotting the relative difference between the fit to EXFOR data and the fit to ENDF/B-VIII.0 evaluated data,
\begin{equation}
\label{eqn:relDiff}
    R_\mathrm{FPY} = \frac{\mathrm{FPY}_\mathrm{EXFOR} - \mathrm{FPY}_\mathrm{ENDF}}{0.5(\mathrm{FPY}_\mathrm{EXFOR} + \mathrm{FPY}_\mathrm{ENDF})}.
\end{equation}
It is unsurprising that the largest differences are seen in the smallest CFYs, i.e., in the symmetric mass region and the wings, where there's little if any experimental data included in the EXFOR fit. 

\subsection{Consistency with Other Fission Observables}
\label{sec:observables}

In this section, we compare our \beoh{} calculations for other prompt and delayed observables to available experimental data.  Although we only included CFYs and $\overline{\nu}_p$ in our optimization, we retain good agreement with other observables of interest.  

\begin{figure*}
    \centering
    \includegraphics[width=\textwidth]{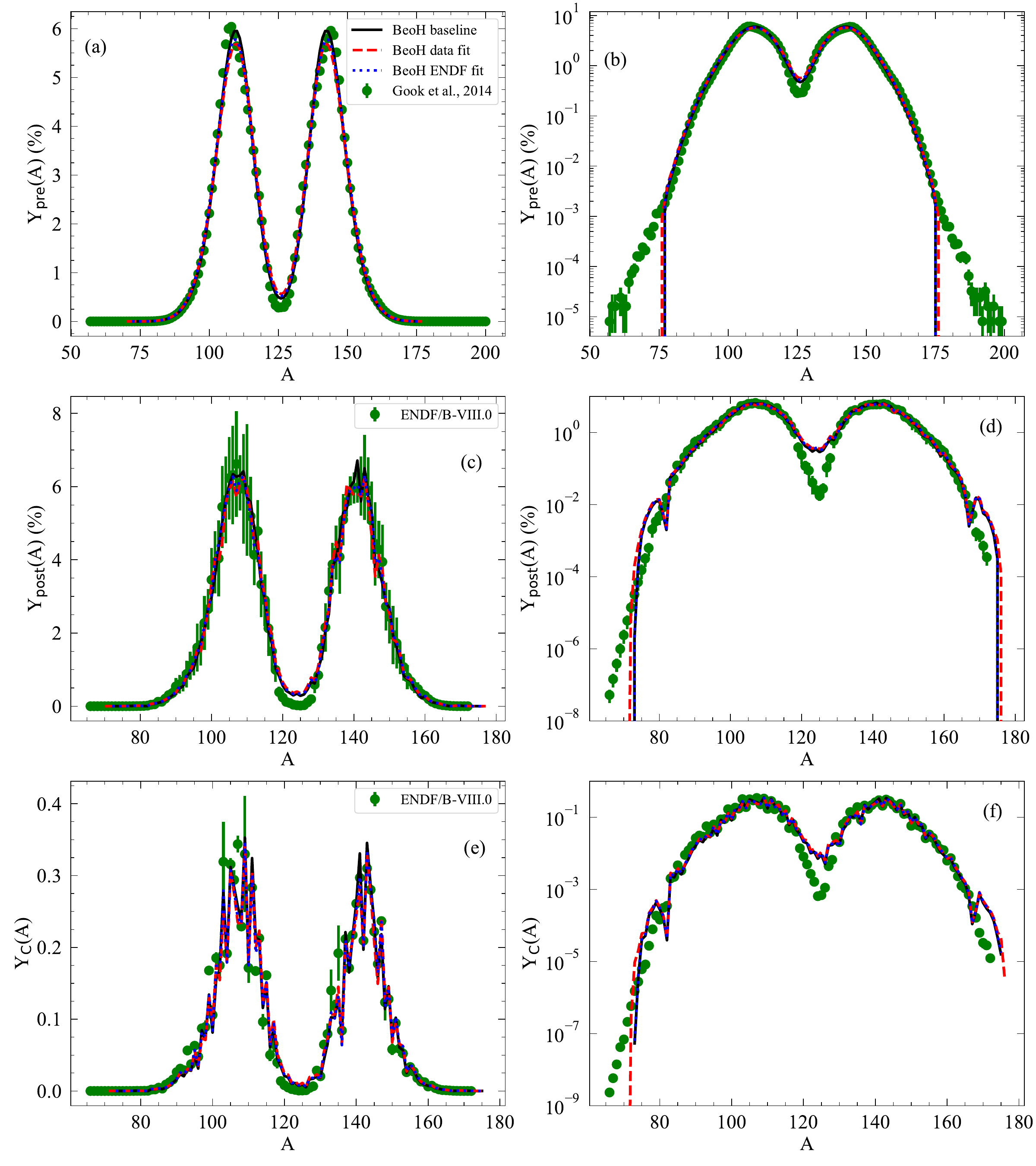}
    \caption{Comparison among experimental data or the ENDF/B-VIII.0 evaluation (green filled circles), the \beoh{} baseline calculation (black solid line), fit to experimental FPYs (red dashed), and fit to ENDF/B-VIII.0 FPYs (blue dashed line) for (a) pre-neutron-emission mass distributions, (b) mass distribution after neutron emission, and (c) mass distribution of cumulative FPYs.}
    \label{fig:YAcomp}
\end{figure*}

Figure \ref{fig:YAcomp} shows a comparison with fission fragment mass distributions of the baseline \beoh{} calculation, the two fits to experimental FPY and ENDF FPY data, along with some selected experimental data \cite{Gook2014} or ENDF/B-VIII.0 evaluation \cite{ENDFB8}.  We show the mass distributions before prompt neutron emission in panel (a), after prompt neutron emission in panel (b), and the cumulative mass distributions in panel (c).  We see that our baseline \beoh{} calculations are higher in the peaks than the calculations that have been fit to both experimental data and ENDF.  We do not reproduce the valley in the symmetric region for any of the comparisons and would need to lower the distribution in other mass ranges due to the sum constraints of the distributions before and after neutron emission, although this should be a small contribution.  Otherwise, for the independent and cumulative mass distributions, we see relatively good agreement with the structure seen in the evaluation.  

\begin{figure}
    \centering
    \includegraphics[width=\columnwidth]{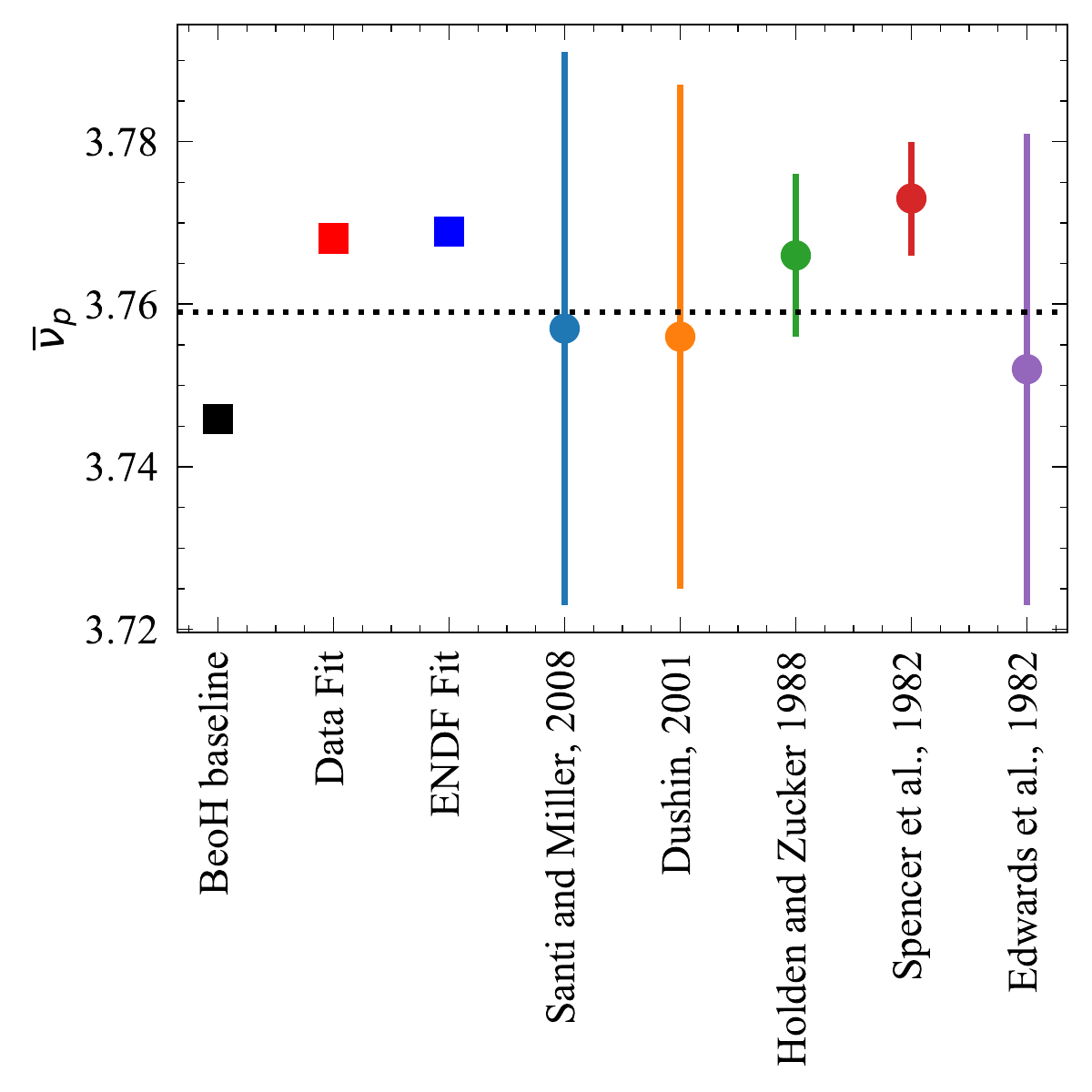}
    \caption{Comparison for average prompt neutron multiplicity of the baseline \beoh{} calculation (black square), \beoh{} fit to experimental data (red square), \beoh{} fit to the ENDF evaluation (blue square), and a subset of available experimental data (filled circles).  ENDF/B-VIII.0 evaluated mean value is shown as the black horizontal dashed line.}
    \label{fig:nupComp}
\end{figure}

We then can compare average multiplicities from \beoh{} and available experimental data.  Figure \ref{fig:nupComp} shows this comparison of the \beoh{} baseline calculation, optimization to experimental data and ENDF evaluation, and selected experimental data \cite{Santi2008,Dushin2004,Holden1988,Spencer1982,Edwards1982} for the average prompt neutron multiplicity, $\overline{\nu}_p$.  The evaluated value from ENDF/B-VIII.0 is shown as the black horizontal line.  Although our fitted values are slightly higher than the evaluated value, it is reasonable within the uncertainties of the experimental data that we included in our fit.  We anticipate that with a more curated set of experimental $\overline{\nu}_p$, we could produce evaluation quality $\overline{\nu}_p$ in addition to the independent and cumulative fission products.  

\begin{figure}
    \centering
    \includegraphics[width=\columnwidth]{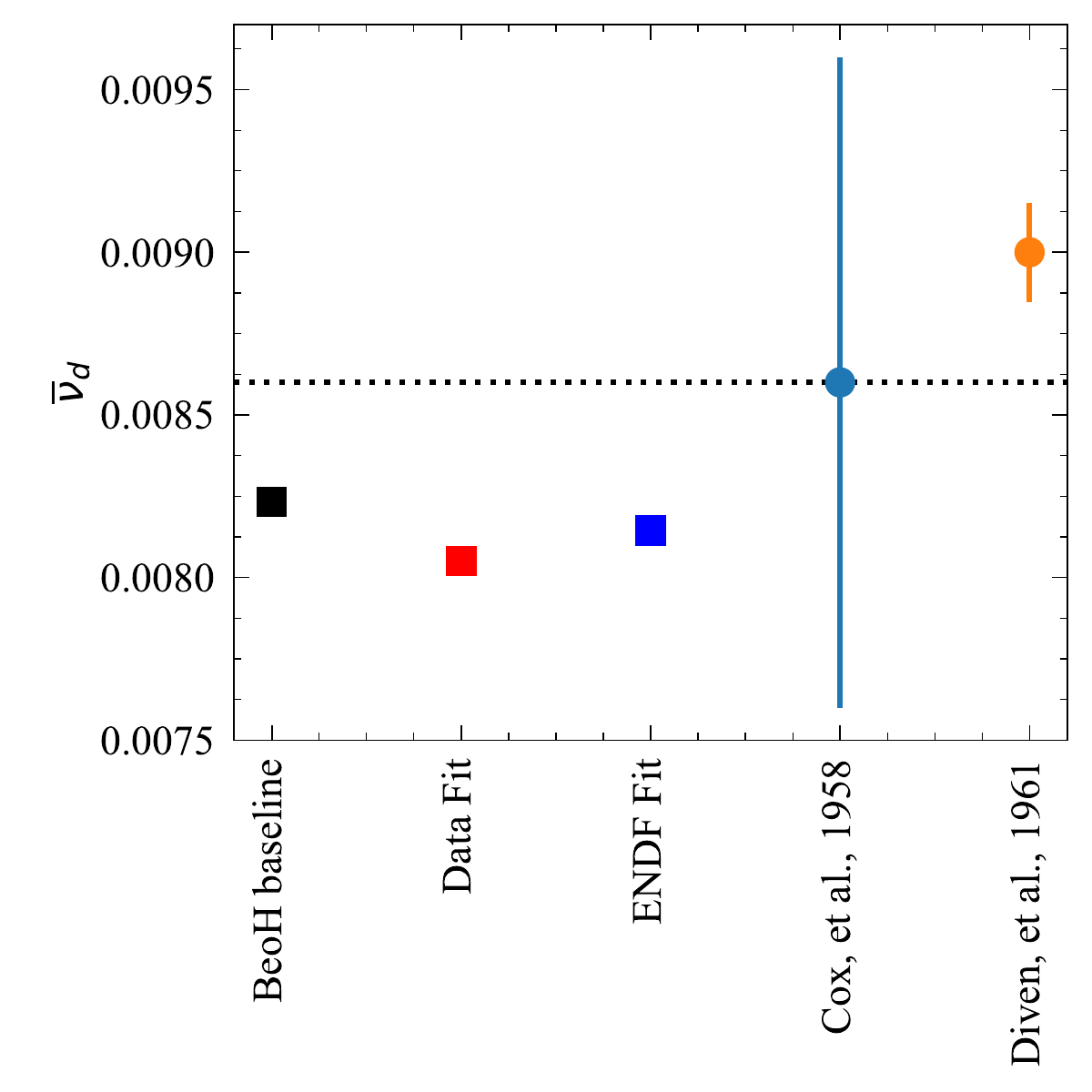}
    \caption{Same as Fig. \ref{fig:nupComp} for the average delayed neutron multiplicity.}
    \label{fig:nudComp}
\end{figure}

Figure \ref{fig:nudComp} compares the average delayed neutron multiplicity, $\overline{\nu}_d$, similar to Fig. \ref{fig:nupComp}, where the data are taken from \cite{Cox1958,Diven1962}.  Considering that this observable was not included in the optimization, we have relatively good agreement with the experimental and evaluated data, particularly the data from Cox \emph{et al.} which is likely the basis of the evaluation.  In the future, we can include $\overline{\nu}_d$ in the optimization procedure which has been shown to impact the odd-even staggering of the charge distribution \cite{Okumura2022}.

\begin{figure}
    \centering
    \includegraphics[width=\columnwidth]{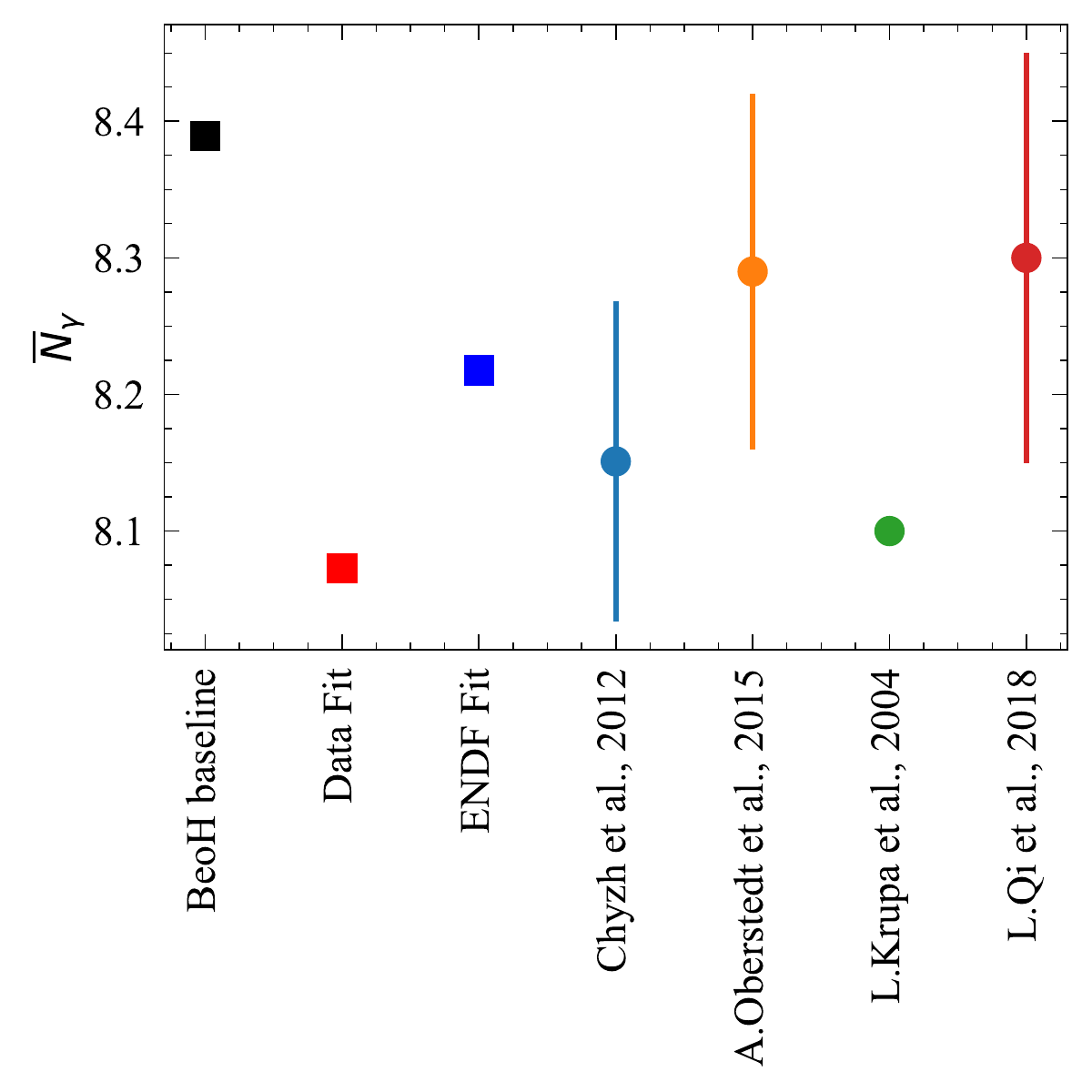}
    \caption{Same as Fig. \ref{fig:nupComp} for the average prompt $\gamma$-ray multiplicity.  The comparison to ENDF is not shown.}
    \label{fig:nugComp}
\end{figure}

Finally, we show a comparison between \beoh{} and select experimental average prompt $\gamma$-ray multiplicity data in Fig. \ref{fig:nugComp}.  We expect some discrepancy between the calculations and experimental data due to the different timing and energy cuts that would have been applied during the experimental measurements due to the detectors used.  These have not been corrected for in the comparison in Fig. \ref{fig:nugComp}, which would increase the experimental $\overline{N}_\gamma$.  Here, we also see a larger discrepancy between the two \beoh{} fits, compared to these differences in the prompt and delayed neutron multiplicities.  The $\gamma$-ray properties of fission reactions are particularly sensitive to the fission fragment spin distribution, \emph{e.g.} \cite{Stetcu2014}.

\section{COVARIANCES}
\label{sec:covariance}

In addition to mean values and standard deviations, the Kalman filter also provides full covariances of the updated FPYs, as given by Eq. (\ref{eqn:fCov}).  In this section, we discuss the calculated CFY covariances and any differences coming from the two optimizations.  The current ENDF/B-VIII.0 library does not contain covariances for the independent or cumulative FPYs, nor does any other international FPY sublibrary.  There have been other efforts to develop covariances for FPYs, such as \cite{Matthews2021,Cheikh2024}, but to date, only JEFF-4.0 has released covariances for some fission reactions at the thermal energy point only.  Here, we present the covariances that accompany the two optimizations discussed in the previous section.

\begin{figure}
    \centering
    \includegraphics[width=\columnwidth]{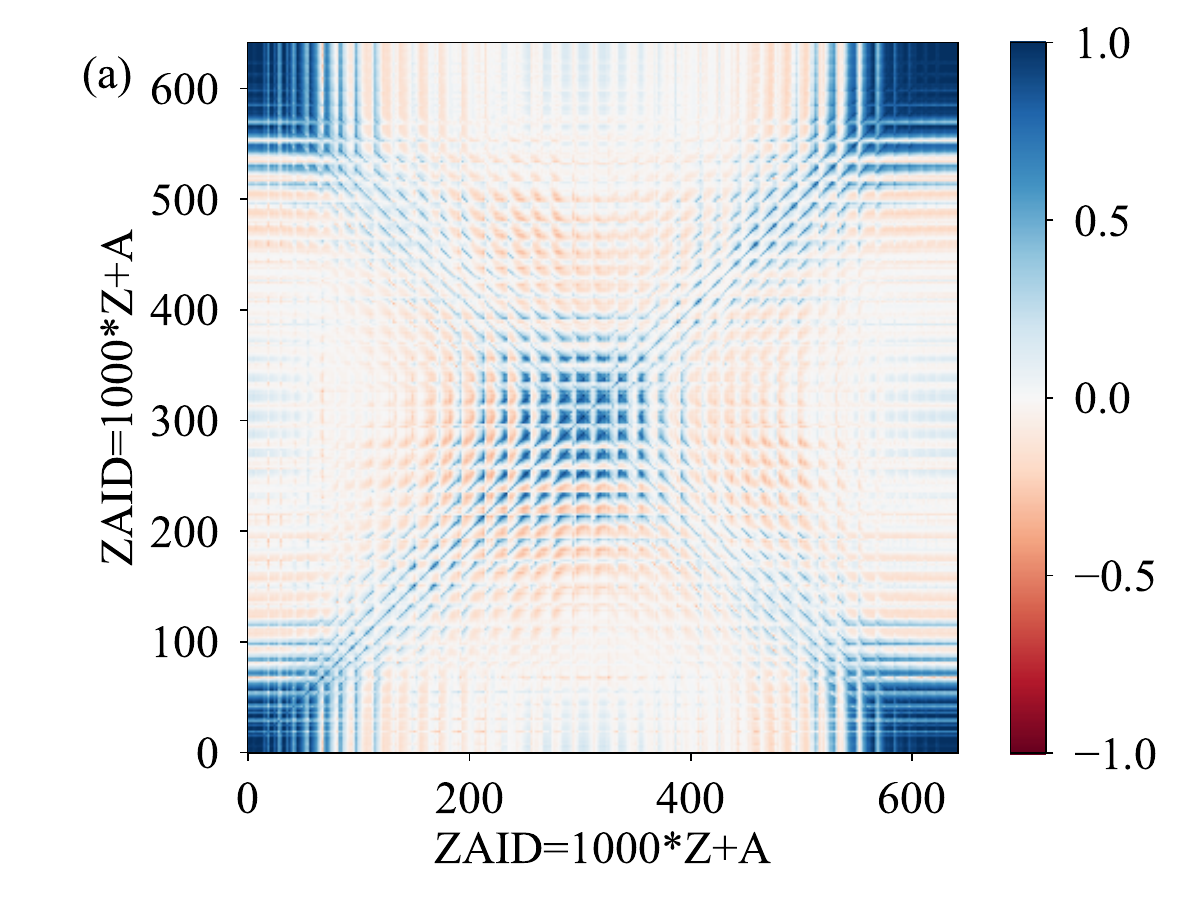} \\
    \includegraphics[width=\columnwidth]{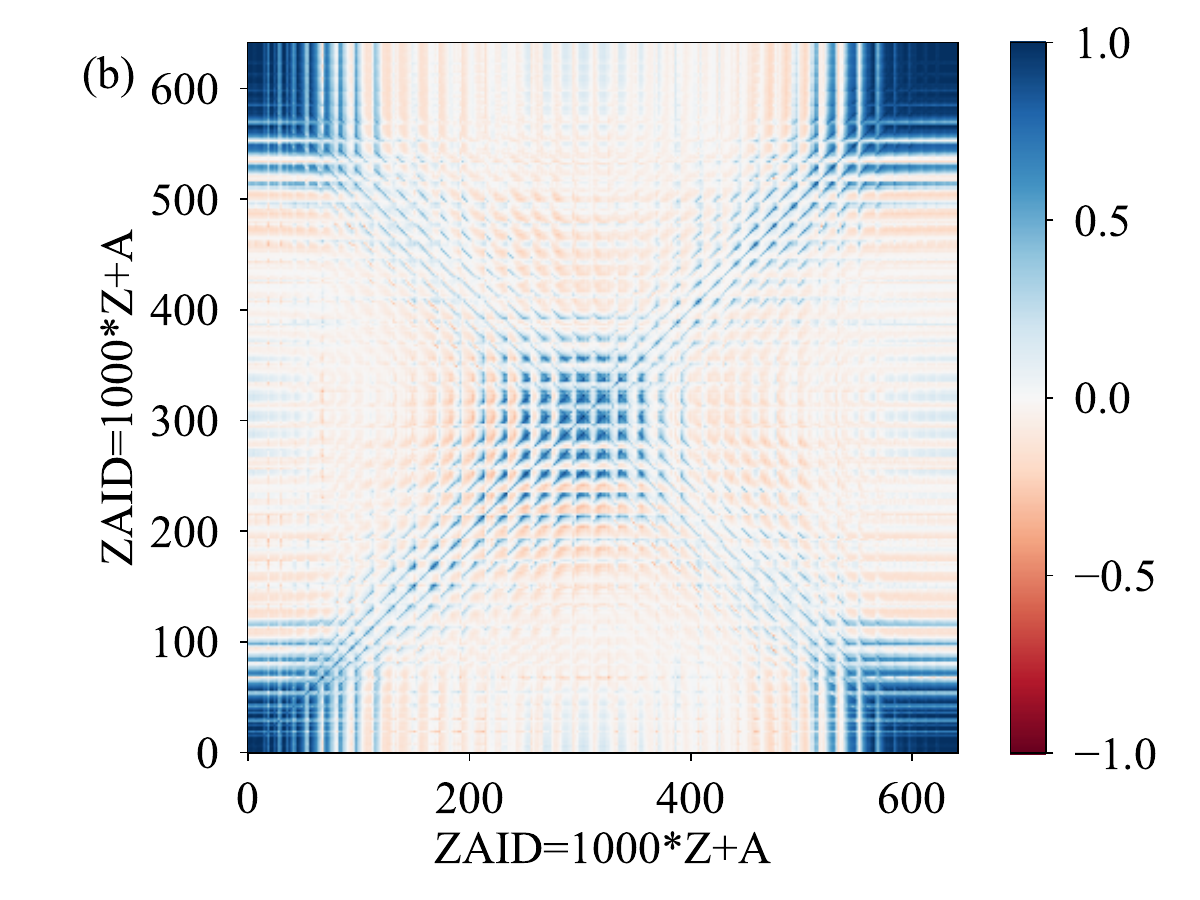} \\
    \caption{Correlations among the fission product yields for the optimization to (a) EXFOR experimental data and (b) ENDF/B-VIII.0 evaluated data.  The indices along the x- and y-axes correspond to approximately increasing ZAID ($1000Z+A$) but do not correspond to the ZAID itself.}
    \label{fig:corrComp}
\end{figure}

Figure \ref{fig:corrComp} shows the correlation matrices for the optimization to EXFOR data and the ENDF/B-VIII.0 evaluation, between the ground state nuclei (isomeric states are not included in these figures).  There are similar features in both correlation plots of Fig. \ref{fig:corrComp}, where the correlations are stronger closer to the diagonal and weaker for the more off-diagonal elements.  We expect some strong correlations based on the physics constraints built into the model.    

\begin{figure}
    \centering
    \includegraphics[width=\columnwidth]{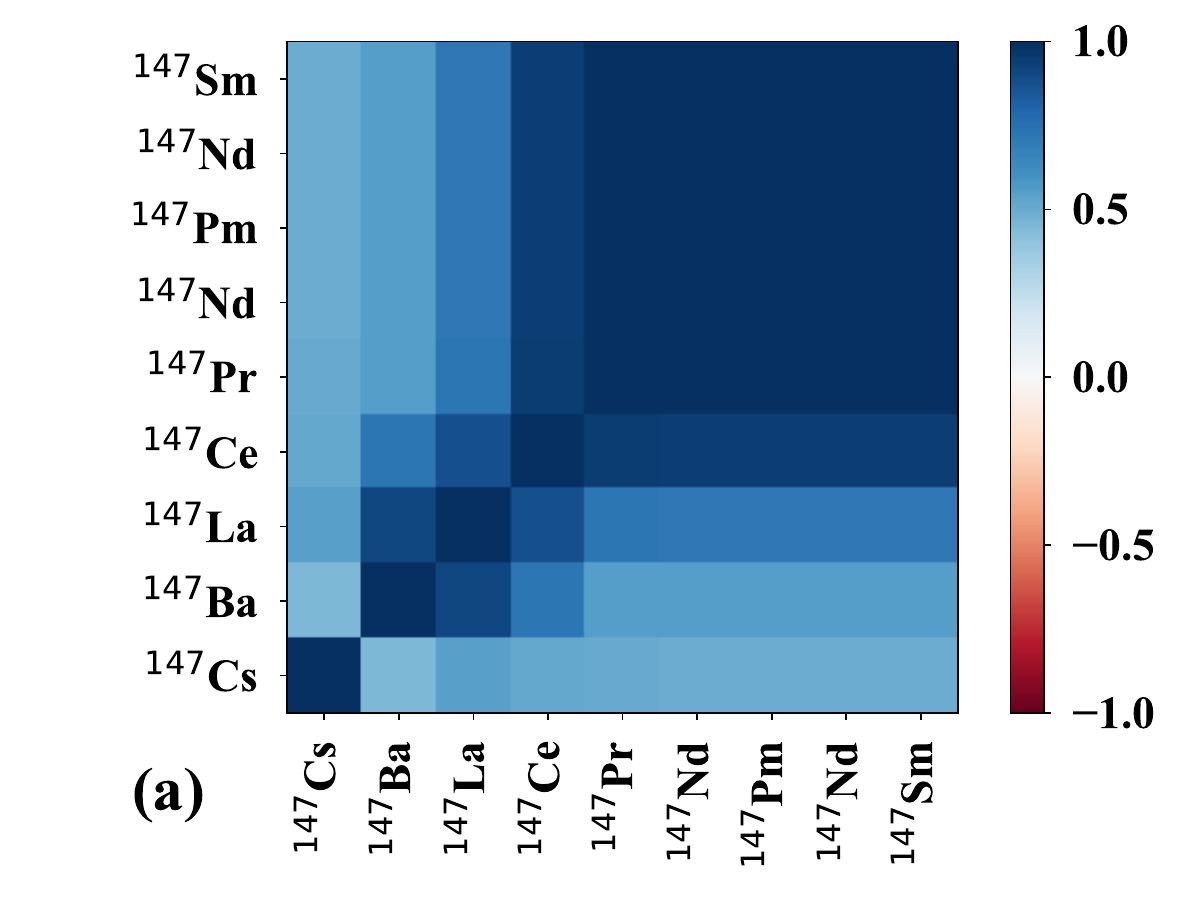} \\
     \includegraphics[width=\columnwidth]{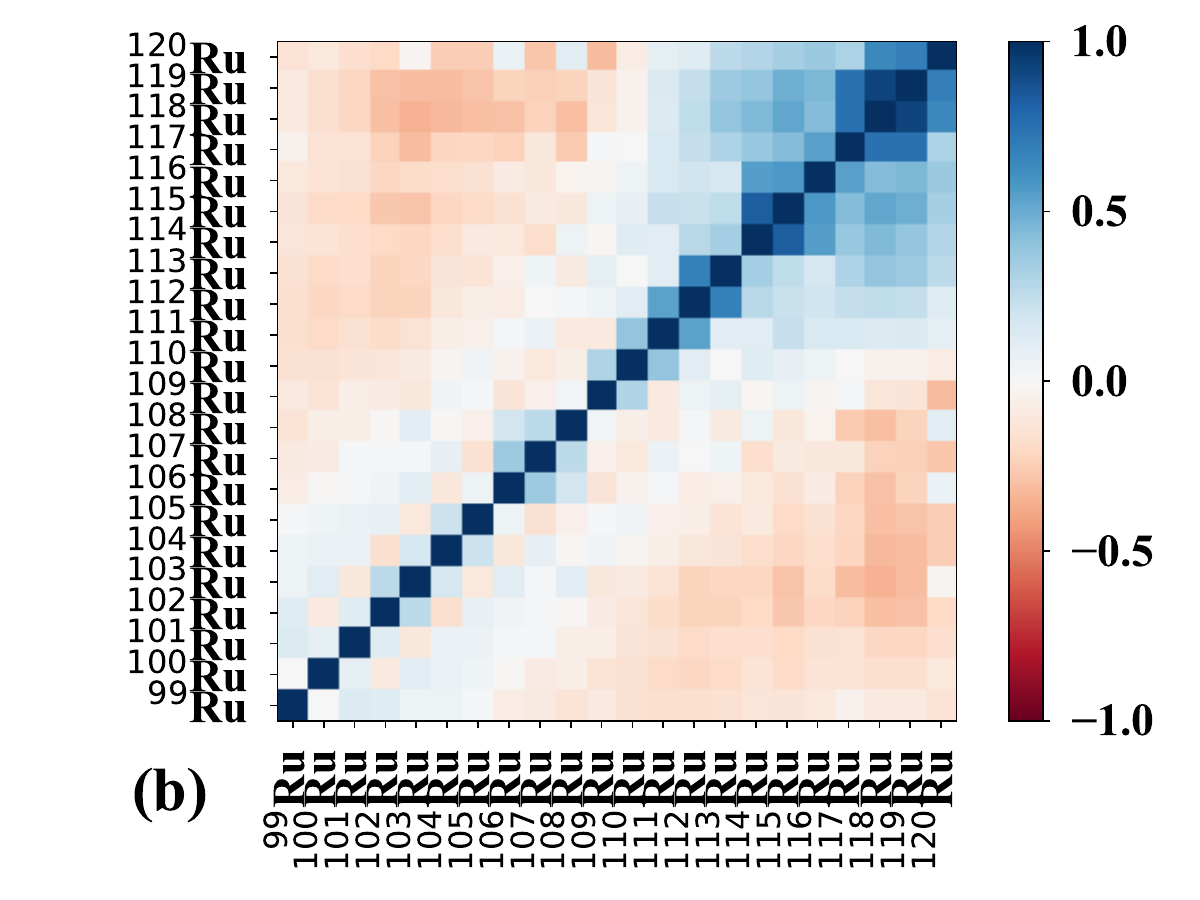}
    \caption{Correlations among nuclei in the (a) $A=147$ mass chain and (b) the Ru ($Z=44$) isotope chain when experimental data from EXFOR were used in the optimization.}
    \label{fig:AZcorr}
\end{figure}

In Fig. \ref{fig:AZcorr}, we investigate these correlations more closely by plotting the correlations among (a) the $A=147$ mass chain and (b) the Ru ($Z=44$) isotopic chain.  Although we just show one representative mass and isotopic chain, the same features are seen in other chains, when there are a sufficient number of nuclei in the chain.  We see that the masses are highly positively correlated.  The correlations among the isotopic chain are much more structured.  There are some weak positive correlations for masses within a few units of one another and weak negative correlations almost everywhere else.  The only exception is in the heaviest mass region where the Ru isotopes are more strongly correlated.  These strongest correlations are seen for isotopic chains where the mass region falls in the symmetric region or the wings of the mass distribution where the FPYs are the smallest.

\begin{figure}
    \centering
    \includegraphics[width=0.97\columnwidth]{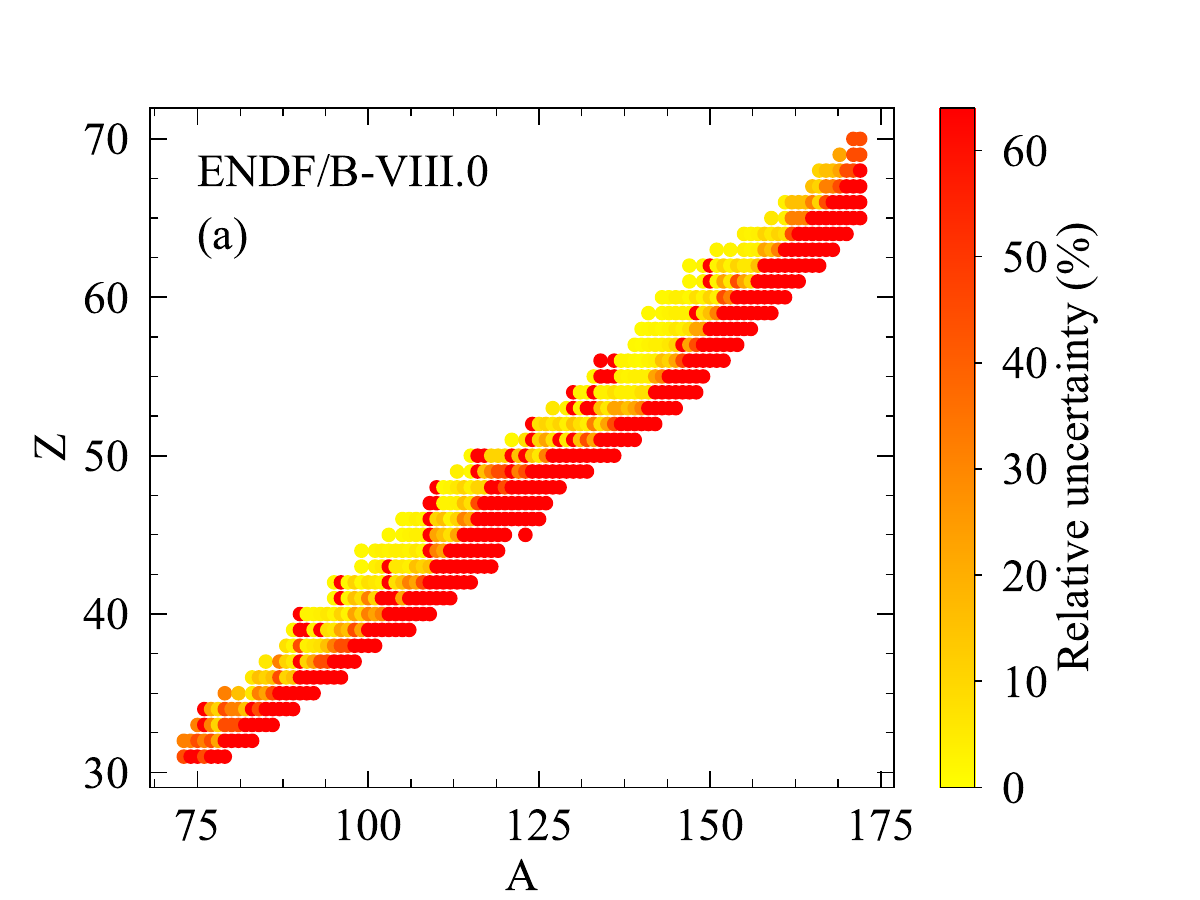} \\
    \includegraphics[width=0.97\columnwidth]{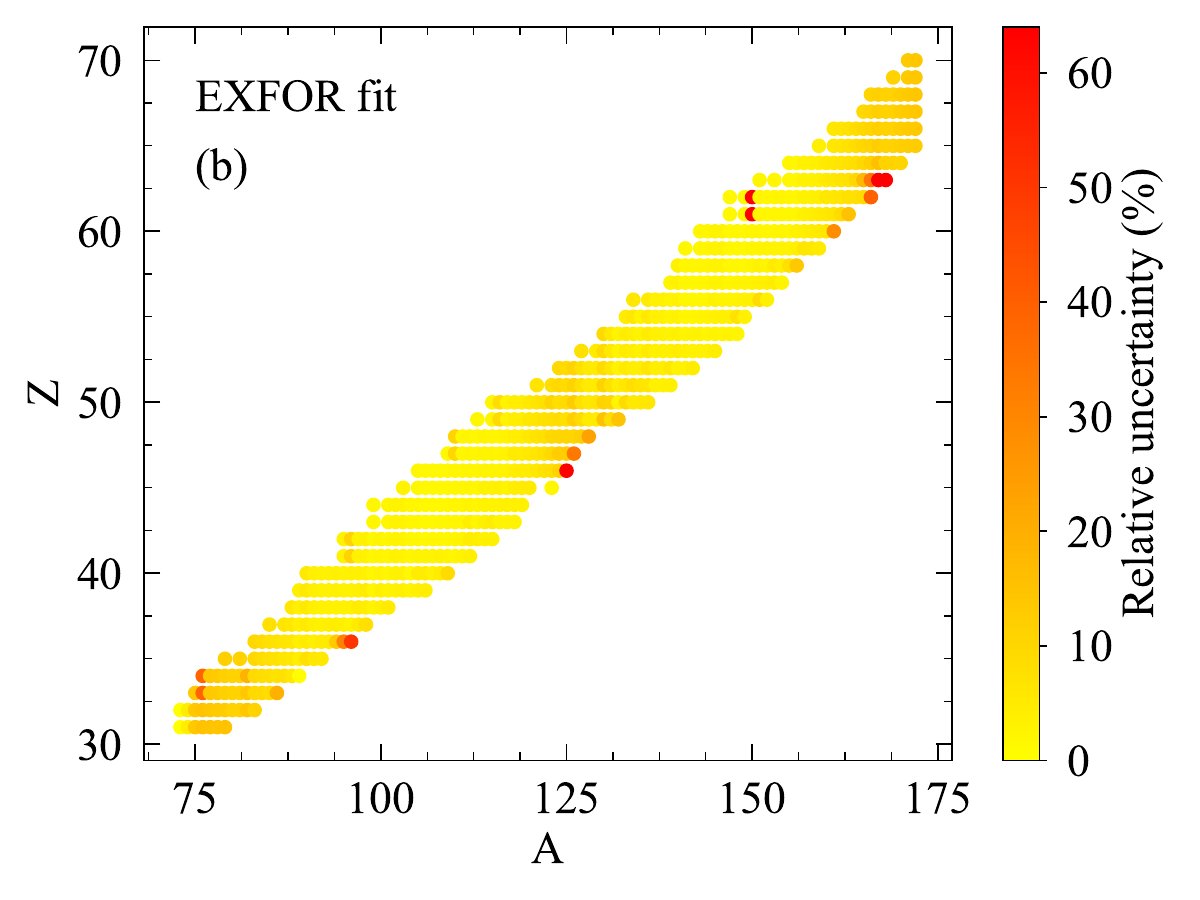} \\
    \includegraphics[width=0.97\columnwidth]{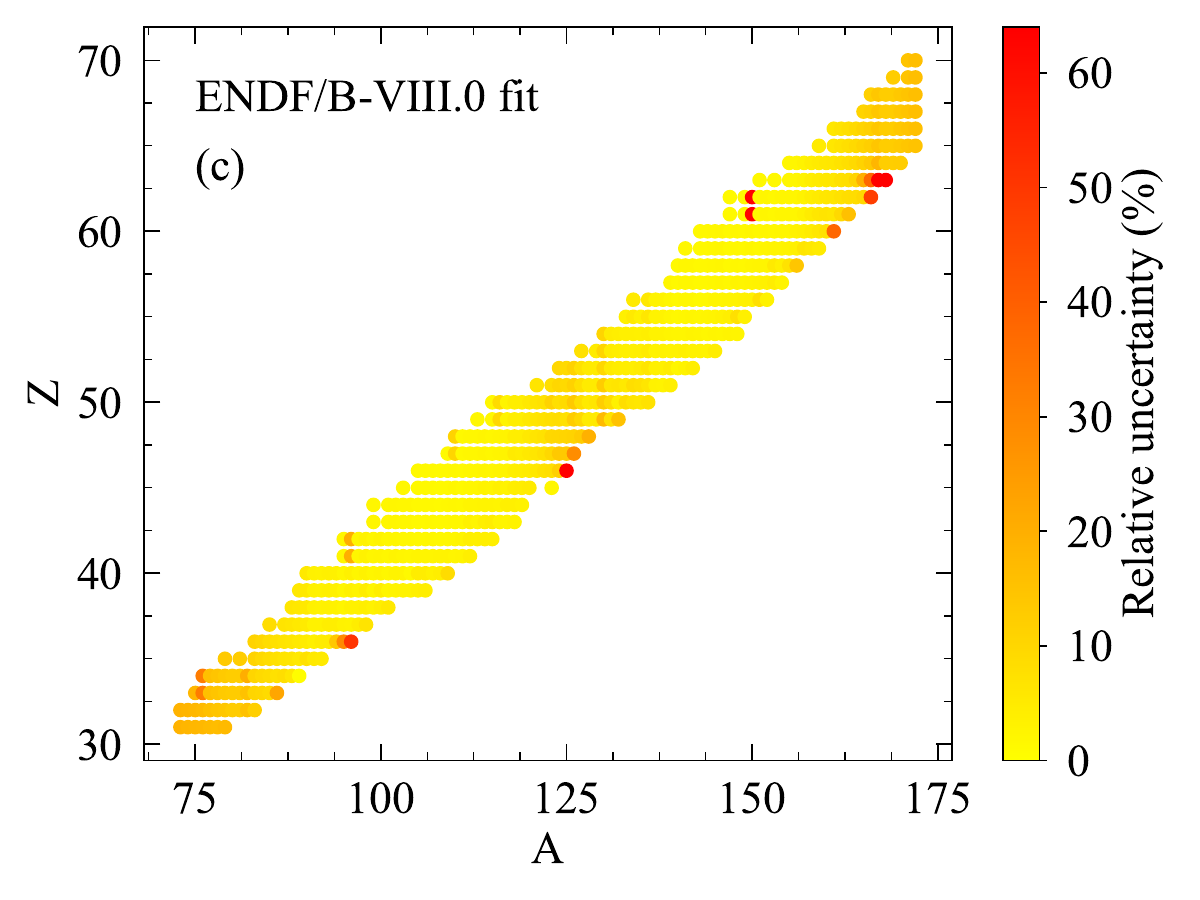} \\
    \caption{Relative uncertainties (\%) of the ground state cumulative FPYs for (a) the current ENDF/B-VIII.0 evaluation, (b) the present fit to EXFOR experimental data, and (c) the present fit to the ENDF/B-VIII.0 evaluation.}
    \label{fig:uncertainties}
\end{figure}

Finally, we plot the relative uncertainties for the ground state cumulative FPYs in Fig. \ref{fig:uncertainties} for (a) the current ENDF/B-VIII.0 evaluation, (b) the present work of fitting to experimental data from EXFOR, and (c) the present work fitting to the ENDF/B-VIII.0 evaluation.  Due to the strong model correlations, the uncertainties on the present FPYs are reduced almost everywhere across the chart of the nuclei, particularly in the highest mass regions.  It is well known that uncertainties produced by a Kalman filter are typically smaller than the spread in data and often have to be increased to be realistic.  Although in this work the covariances from Kalman are increased by the $\chi^2/N$ of the optimization, likely the uncertainties should be further increased to reflect the true uncertainty of the data.  Still, the large decrease in uncertainty in the high-mass region would persist.  

\section{CONCLUSIONS}
\label{sec:conclusions}


In summary, we have detailed a path towards an updated procedure for evaluating independent and cumulative fission product yields by combining experimental data and a Hauser-Feshbach fission fragment decay model.  This method has been applied to the spontaneous fission of $^{252}$Cf.  We use \beoh{}, the deterministic LANL-developed Hauser-Feshbach fission fragment decay code, to calculate cumulative fission product yields consistently with other delayed fission observables and prompt fission observables, such as delayed and prompt neutron multiplicity, prompt $\gamma$-ray multiplicity, multiplicity distributions, energy spectra, and independent fission product yields.  The model calculations are combined with experimental data for the cumulative fission product yields, prompt and delayed neutron multiplicities through a Kalman filter optimization to produce optimized parameter values in conjunction with covariances.    

We compare optimizations using experimental data from the EXFOR database and evaluated values from the ENDF/B-VIII.0 library.  While there are some differences between cumulative fission product yield mean values when the two databases are used for the optimization--particularly for nuclei where there are no experimental data--there is fairly good agreement between the two optimizations.  Although the uncertainties are significantly different for many nuclei, the features seen in the correlation matrices are consistent between the two optimizations meaning that they are largely due to correlations from the model and not introduced by the choice of data.

Similar evaluations for neutron-induced fission of $^{235}$U, $^{238}$U, and $^{239}$Pu have been performed and validation studies are underway.  Additional challenges come with the inclusion of the energy dependence in neutron-induced fission.  More model parameters are included in the optimization, not only because of this energy dependence but also because of the opening of the multi-chance fission channels, where multiple compound nuclei can fission.  \beoh{} has already been extended to take into account the energy dependence beyond first-chance fission \cite{Okumura2018,Lovell2021} and preliminary optimizations have already been performed.  Those results will be the subject of a subsequent publication.  

Additionally, with the development of full covariances, it will be important to propagate these covariances through applications and understand the impact of new uncertainties and correlations.  Such studies are already being performed for the neutron-induced fission product yields.

\acknowledgments

This work was performed under the auspice of the U.S. Department of Energy by Los Alamos National Laboratory under Contract 89233218CNA000001 and was supported by the Office of Defense Nuclear Nonproliferation Research and Development (DNN R\&D), National Nuclear Security Administration, U.S. Department of Energy.

\bibliography{References}







\end{document}